\documentclass[11pt]{article}
\pdfoutput=1  
\usepackage{graphicx}
\graphicspath{{./figures/}}
\usepackage{appendix}
\usepackage{soul}
\usepackage{verbatim}
\usepackage{latexsym,amsmath,amsfonts,amssymb,booktabs}
\usepackage[font=small]{caption}
\usepackage{slashed,upgreek,amscd,cancel,tensor,color}
\usepackage{adjustbox}
\usepackage[numbers,compress,square]{natbib}
\usepackage{epsfig,latexsym}
\usepackage[pdfencoding=auto]{hyperref}
\usepackage{url}
\numberwithin{equation}{section}
\usepackage{doi}
\usepackage{subcaption}
\usepackage[normalem]{ulem}
\definecolor{MyBlue}{rgb}{0.15,0.15,0.70}

\hypersetup{
colorlinks=true,
citecolor=MyBlue,
linkcolor=MyBlue,
urlcolor=MyBlue
}

\input epsf
\usepackage{amssymb}
\usepackage{amsmath}
\usepackage{amsfonts}
\usepackage{upgreek}
\usepackage{latexsym}
\usepackage{stfloats}
\usepackage{afterpage}

\def\re#1{(\ref{#1})}

\newcommand{\be}{\begin{equation}}
\newcommand{\ee}{\end{equation}}
\newcommand{\beq}{\begin{equation}}
\newcommand{\eeq}{\end{equation}}
\newcommand{\bea}{\begin{eqnarray}}
\newcommand{\eea}{\end{eqnarray}}

\newcommand{\HH}{{\cal H}}

\newcommand{\m}{{\rm m}}

\newcommand{\bfp}{{\bf p}}

\usepackage[stable]{footmisc}

\def\dkmu2{\delta K_{\mu \nu}\delta K^{\mu \nu}}
\def\pmu2{  \phi_{\mu \nu}\phi^{\mu \nu}}

\newcommand{\bfQ}{{\bf{Q}}}

\newcommand{\Ic}{{\cal I}}

\newcommand\bk{{\mathbf{k}}}

\newcommand{\vphi}{\varphi}

\newcommand{\bv}{\bf v}
\newcommand{\bx}{{\bf x}}
\newcommand{\bq}{\mathbf q}
\newcommand{\bd}{\mathbf d}

\newcommand\ees{\end{eqnarray}}
\newcommand\bees{\begin{eqnarray}}

\newcommand{\cH}{\mathcal{H}}

\newcommand{\Fn }{F_{n}}
\newcommand{\Fone}{F_{1}}
\newcommand{\Ftwo}{F_{2}}
\newcommand{\Fthree}{F_{3}}
\newcommand{\Gn }{G_{n}}
\newcommand{\Gl }{G_{l}}
\newcommand{\Gone}{G_{1}}
\newcommand{\Gtwo}{G_{2}}
\newcommand{\Gthree}{G_{3}}
\newcommand{\Kn }{K_{n}}
\newcommand{\Knmenom }{K_{n-m}}
\newcommand{\Knmenol }{K_{n-l}}
\newcommand{\Kone}{K_{1}}
\newcommand{\Ktwo}{K_{2}}
\newcommand{\Kthree}{K_{3}}
\newcommand{\Kfour}{K_{4}}
\newcommand{\AC}{a}
\newcommand{\BC}{d}
\newcommand{\CC}{c}
\newcommand{\hh}{h}

\newcommand{\egi}{EGI}

\begin{document}
\vspace{0.5cm}

\begin{center}
\Large{\textbf{The Large Scale Structure Bootstrap:\\
perturbation theory and bias expansion from symmetries
}} 

\vspace{.2cm}
G.~D'Amico$^{1}$, M.~Marinucci$^{1}$, M.~Pietroni$^{1}$, and F.~Vernizzi$^{2}$
\\ \small{
\textit{$^1$ Department of Mathematical, Physical and Computer Sciences, University of Parma, and INFN Gruppo Collegato di Parma, Parco Area delle Scienze 7/A, 43124, Parma, Italy\\
$^2$ Institut de physique th\' eorique, Universit\'e  Paris Saclay \\ [0.05cm] CEA, CNRS, 91191 Gif-sur-Yvette, France  }}

\vspace{.2cm}

\vspace{0.5cm}
\today

\end{center}


\begin{abstract}

 We investigate the  role played by symmetries in the perturbative expansion of  the large-scale structure. In particular, we establish which of the coefficients of the perturbation theory kernels are dictated by symmetries and which not. Up to third order in perturbations, for the dark matter density contrast (and for the dark matter velocity)   only three coefficients are not fixed by symmetries and  depend on the particular cosmology. For generic biased tracers, where number/mass and momentum conservation cannot be imposed in general, this number rises to seven  in agreement with other bias expansions discussed in the literature.  A crucial role in our analysis is provided by extended Galilean invariance, which follows from diffeomorphism invariance in the non-relativistic limit. We identify a full hierarchy of extended Galilean invariance constraints, which fix the analytic structure of the perturbation theory kernels as the sums of an increasing number of external momenta vanish.
Our approach is especially relevant for non-standard models that respect the same symmetries as  $\Lambda$CDM and where perturbation theory at higher orders has not been exhaustively explored, such as   dark energy and modified gravity scenarios. 
In this context, our results can be used to systematically extend the bias expansion to higher  orders and  set up model independent analyses.

\end{abstract}


\tableofcontents

\vspace{.5cm}

\section{Introduction}
\label{sec:intro}

Data from next generation Large Scale Structure (LSS) surveys (such as DESI \cite{DESI:2016fyo} and Euclid \cite{Amendola:2016saw}) are expected to shed light on the most fundamental open questions facing cosmology and particle  physics; among them, the absolute scale of neutrino masses, the origin of the cosmic acceleration, and the properties of the initial conditions of cosmic perturbations generated during Inflation.  

The main obstacle in extracting information from data lies in the nonlinear nature of the evolution of structures, which becomes more and more relevant the smaller the scales and the lower the redshift. Moreover, the process of galaxy formation is not completely understood. Different and complementary approaches have been developed to deal with this issue and can now be employed in synergy.

N-body simulations provide the most accurate description of cold dark matter dynamics, but computational times limit the range of scales and of parameter space that can be directly simulated.
To simulate directly the galaxies (which is what we observe), one would need hydrodynamic codes, which however are less reliable.
In order to enlarge the available portion of parameter space, cosmological {\it emulators} are being developed (see, for instance, \cite{Lawrence:2009uk, Nishimichi:2018etk, Euclid:2018mlb}), which provide fast interpolation tools, calibrated with a large suite of N-body simulations run for different points in parameter space.  

A complementary tool is provided by analytical approaches, which, in recent years, have reached a stage of maturity.
Different formulations \cite{PT,RPTa,MP07b,Taruya2007,Pietroni08,Bernardeau:2008fa,Baumann:2010tm,Pietroni:2011iz,Carrasco:2012cv,Manzotti:2014loa,Senatore:2014via,Baldauf:2015xfa,Blas:2016sfa,Noda:2017tfh,Ivanov:2018gjr} have by now converged to a successful description based on effective theory ideas,
 at least for what concerns the galaxy Power Spectrum (PS).
The first ingredient of the analytic framework is perturbation theory (PT) (for reviews, see \cite{PT,Bernardeau:2013oda}). The nonlinear fields on large scales are  expressed as an expansion in terms of the initial (linear) fields, convolved in momentum space via kernels that can be derived, in a given model, by iteratively solving the evolution equations.
Because the pressureless perfect fluid approximation breaks down at short scales, standard perturbation theory is not guaranteed to converge to the correct answer. Therefore, one should renormalize the theory by appropriate UV counterterms, organized in an expansion in powers of momentum. They describe the effect of short scales on longer, perturbative ones \cite{Baumann:2010tm,Pietroni:2011iz,Carrasco:2012cv}. Finally, the observed galaxy field is connected to the computed dark matter one by means of a {\it bias expansion},  which is an expansion in derivatives of the gravitational potential and the velocity field (for a review, see \cite{Desjacques:2016bnm}).
The practical applicability of the effective theory approach has been greatly enhanced  by Fast Fourier Transform methods to evaluate the momentum integrals appearing in the computation of the PS beyond linear order \cite{Schmittfull:2016jsw,McEwen:2016fjn,Simonovic:2017mhp}, which allow an efficient implementation of Markov Chain Monte Carlo algorithms for parameter estimation. 
The perturbative approach has been successfully applied to current data to extract cosmological parameters \cite{DAmico:2019fhj, Ivanov:2019pdj,Colas:2019ret}.
An alternative, more recent, approach with respect to computing correlators consists in considering  the full theoretical galaxy field and comparing it  with observations, after marginalizing on the phases of the initial conditions \cite{Schmidt:2018bkr}. Also in this case, a theoretical model for the evolution of the field is required, and PT based approaches are one option. 

In this paper we investigate the role of symmetries in dictating the structure of  the PT kernels, independently on the details of the equations of motion. This is particularly relevant {when considering models beyond $\Lambda$CDM, such as  for dark energy and modified gravity.  In particular, we will show that the structures of the kernels in all dark energy and modified gravity models with the same symmetries as $\Lambda$CDM are the same. Only a few time-dependent coefficients change, depending on the particular  expansion history and possible time-dependence of the modified gravity parameters.

One  symmetry plays a crucial role in  fixing the structure of the kernels both for dark matter (DM) and for generic biased tracers:  our results are based on the assumption that the dynamics is invariant under  uniform spatial translations with arbitrary time dependence, which follows from diffeomorphism invariance in the non-relativistic limit \cite{Creminelli:2013mca}, the regime considered here.  In the following we will refer to this  as {\em extended} Galilean invariance (\egi) \cite{Jain:1995kx,Scoccimarro:1995if} to distinguish it from the usual Galilean invariance relating inertial frames by a boost. Here this requirement translates in constraints on the analytic structure of the PT kernels. 

We  identify  a full hierarchy of EGI constraints, depending on the PT order of the induced displacement field. 
If it is of linear order, the corresponding `Leading Order' (LO) constraints, given in Eq.~\re{GIc},  fix the pole structure of the kernels as one of its external momenta vanishes. If it is of second PT order, we get `Next-to-Leading' (NLO) constraints on the kernels as the sum of two external momenta vanishes (see Eq.~\re{nlo}), and so on. Up to third PT order, only constraints up to NLO are relevant, but we derive the general formula for the EGI constraints, that can be applied at any higher PT order, given in Eq.~\re{genGIshort}.

The difference between DM and  biased tracers is just mass and momentum conservation, which holds for the former but not, in general, for the latter. As a consequence,  DM kernels, for both density and velocity, are more constrained than generic tracer kernels. Up to third PT order, they contain just three model-dependent coefficients. 
On the other hand, in the case of biased tracers we end up  with seven independent coefficients, equivalently to the other bias expansions discussed in the literature for fixed cosmologies \cite{Chan:2012jj,Saito:2014qha,Assassi:2014fva,Mirbabayi:2014zca,Senatore:2014eva,Angulo:2015eqa,Fujita:2016dne,Eggemeier:2018qae}.

Our approach, exploiting all the possible information dictated by symmetries without considering explicit equations of motions, is analogous in spirit to the one of the `conformal bootstrap' \cite{Ferrara:1973yt,Polyakov:1974gs}, which is experiencing a revival in many areas of modern theoretical physics, from field theory (for a review, see \cite{Simmons-Duffin:2016gjk})  to inflationary cosmology \cite{Arkani-Hamed:2018kmz,Pajer:2020wnj}.  More specifically, for large-scale structure, a first step in this direction was taken in \cite{Mercolli:2013bsa}, where symmetries were used to derive the fluid equations of motion. Another relevant reference is \cite{Fujita:2020xtd}, where the consistency relations  for the LSS \cite{Peloso:2013zw,Kehagias:2013yd} were imposed on equal-time field correlators to derive constraints on the PT kernels. As is well known, consistency relations are based on the same symmetry considered here, namely \egi, plus a series of assumptions on the `state', namely, adiabatic and Gaussian initial conditions. The results of our approach, which operates at the field level, are compatible with those of \cite{Fujita:2020xtd}, but we are able to fix the two `universal' coefficients which were left free in \cite{Fujita:2020xtd} (see sect.~\ref{sectbias} for a detailed discussion on this point).

The approach presented in this paper provides a useful tool in two respects. On one hand, the deep role played by EGI, revealed by our analysis, provides a systematic way to derive the structure of the PT kernels  for biased tracers at orders beyond the third,  when NNLO, and so on, constraints start to play a role. On the other hand, the framework presented here is optimal if one is interested in model-independent analyses of beyond $\Lambda$CDM scenarios,  since, as we will see, we will be able to single out coefficients which depend only on cosmology and not on the tracer type. 

The paper is organized as follows. In sect.~\ref{constr} we introduce the PT kernels which will be the subject of our analysis and derive the constraints coming from EGI, both at LO and at a generic order, and mass and momentum conservation. In sect.~\ref{matterk} we impose all the above constraints, to derive the most general structure for the DM density and velocity kernels, up to third order. We also show how the arbitrary coefficients can be fixed by assuming a definite cosmological model, and discuss explicitly the case of $\Lambda$CDM and of the nDGP model \cite{Dvali:2000xg}. In sect.~\ref{gentrac} we relax the constraints from mass and momentum conservation, to derive the kernels for biased tracers. We also discuss the relation between our results and other bias expansions derived in the literature.  In sect.~\ref{sec:HD} we discuss the corrections to the kernels necessary in order to improve the description of the effects of short (UV) modes. Finally, in sect.~\ref{concl},  we  summarize and conclude. 

Moreover, in Appendix~\ref{A1}, we give a more general derivation of the EGI constraints with respect to the one discussed in the text for sake of clarity, in Appendix~\ref{nnlosec} we work out in detail the EGI constraint at the NNLO order, in Appendix~\ref{app:kernels} we give details on the derivation of the kernels and, finally, in Appendix~\ref{ansols} we give the analytic solutions of the equations for the time-dependent coefficients in $\Lambda$CDM and nDGP.

\section{Constraints on PT kernels for general tracers }
\label{constr}

Let us use the logarithm of the scale factor as time variable, $\eta\equiv \log\left({a}/{a_0}\right) $, and  introduce  the linear growth factor   $D(\eta)$ and  the linear growth rate
\be
\label{growth}
f(\eta)\equiv\frac{d \log D(\eta)}{d\eta} \;.
\ee
We denote by $D_+(\eta)$ and $D_-(\eta)$  the growing and decaying mode of  the growth factor, respectively.  For the growth rate, an analogous definition of $f_+$ and $f_-$ follows from eq.~\eqref{growth}.

Beside the matter density contrast $\delta (\bx,\eta)\equiv \rho(\bx,\eta)/\rho_0(t)-1$, where $\rho$ is the energy density and $\rho_0$ its background value, and the matter velocity ${\bf v}(\bx,\eta)$, we define  the rescaled velocity divergence,
\be
\Theta (\bx,\eta) \equiv - \frac{\nabla \cdot {\bf v}(\bx,\eta)}{ f_+(\eta) \cH(\eta)}\;.
\label{vdiv}
\ee 
In the following, we will also consider the number density contrast of any tracer field $ \delta_{t}(\bx,\eta)$, which can be galaxies, halos, etc. Using perturbation theory we can expand these quantities as
\be
\label{expansion}
\delta (\bx,\eta) = \sum_{n=1}^\infty \,\delta^{(n)} (\bx,\eta), \qquad \Theta (\bx,\eta) = \sum_{n=1}^\infty \,\Theta^{(n)} (\bx,\eta), \qquad   \delta_{t} (\bx,\eta) = \sum_{n=1}^\infty \,\delta^{(n)}_t (\bx,\eta)  \;.
\ee

For the linear field it is convenient to introduce the {\em linear} doublet  $\phi^{\lambda}_\bk(\eta)$ ($\lambda=1,2$) that includes  the linear density contrast  and the linear velocity divergence,
\be 
\phi^1(\bx,\eta ) \equiv \delta^{(1)} (\bx, \eta)\,,\qquad \phi^2(\bx,\eta )\equiv  \Theta^{(1)} (\bx,\eta) \,.
\label{linfields}
\ee
After decomposing it in Fourier space,
\be
 \phi^{\lambda} (\bx,\eta)= \int \frac{d^3 k}{(2\pi)^3} e^{- i \bk \cdot \bx } \,\phi^{\lambda}_\bk(\eta)   \;,
\ee
for each mode $\bk$ the  linear solution
reads
\beq
\phi^{\lambda}_\bk(\eta) = u^\lambda_{f}(\eta) \vphi_\bk(\eta) \,, 
\qquad u^\lambda_{f}(\eta)\equiv \left(\begin{array}{c} 1\\ \frac{ f(\eta)}{ f_+(\eta)} \end{array}\right)\,,
\label{philin}
\eeq
where $\vphi_\bk(\eta)$ is related to the initial field $\vphi_\bk(0)$ by the linear growth, $\vphi_\bk(\eta)  = D(\eta) \vphi_\bk(0) $, with $D(0)=1$.
For simplicity, and because this is what happens in most cosmological situations, in the following we will assume that  the linear fields are in the growing mode, in which case $u^\lambda_{f_+}(\eta) \equiv (1,1)^T$. One can straightforwardly generalize our discussion to include the decaying mode as well.

\subsection{Perturbation theory kernels}

For the time being, we will consider only {\em deterministic evolution}, i.e., that the  matter density contrast $\delta(\eta,\bx)$, the  rescaled velocity divergence $ \Theta (\eta,\bx)$, and any tracer $\delta_{t}(\eta,\bx)$, are functionals of the linear matter density field $\varphi (\eta,\bx)$. Later, in sect.~\ref{sec:HD}, we will discuss the stochastic contributions to the field evolution induced by small scale modes.

For a deterministic evolution, the non-linear fields can be written as
 \begin{align}
\label{functional}
 \delta(\bx,\eta) &= {\cal F}[\vphi](\bx,\eta)\, , \qquad  \Theta(\bx,\eta) = {\cal G}[\vphi](\bx,\eta)\, , \qquad \delta_{t}(\bx,\eta) = {\cal K}[\vphi](\bx,\eta)\,. 
 \end{align}
 In general, the non-linear velocity field has also a vorticity component, besides the divergence one. In standard perturbation theory, vorticity decays  as $1/a$ at linear order \cite{PT} and,  setting its initial condition to zero, it is not generated at any higher order. Therefore, as long as deterministic evolution is considered, it is consistent to consider a curl-free velocity field. On the other hand, small-scale stochasticity sources vorticity, as discussed in \cite{Mercolli:2013bsa, Carrasco:2013mua}, and symmetry arguments can be used to constrain  these contributions as well. However, as concluded for instance in \cite{Pueblas:2008uv} the backreaction of vorticity on the density and velocity divergence correlators is extremely suppressed on the scales reachable by PT methods, therefore we will not include these effects in this work.
 
Expanding  the LHS of eq.~\re{functional} in perturbation theory using eq.~\eqref{expansion} in Fourier space and expanding the RHS in the field $\varphi$, we obtain 
\begin{align}
\delta^{(n)}_{\bk}(\eta) &\equiv \Ic_{\bk;\bq_1\cdots,\bq_n}\, \Fn(\bq_1,\cdots,\bq_n;\eta) \vphi_{\bq_1}(\eta)\cdots  \vphi_{\bq_n}(\eta),
\label{expansion_d2} \\
\Theta^{(n)}_{\bk}(\eta) &\equiv \Ic_{\bk;\bq_1\cdots,\bq_n}\, \Gn(\bq_1,\cdots,\bq_n;\eta) \vphi_{\bq_1}(\eta)\cdots  \vphi_{\bq_n}(\eta),
\label{expansion_t2} \\
\delta^{(n)}_{t,\bk}(\eta) &\equiv \Ic_{\bk;\bq_1\cdots,\bq_n}\, \Kn(\bq_1,\cdots,\bq_n;\eta) \vphi_{\bq_1}(\eta)\cdots  \vphi_{\bq_n}(\eta),
\label{expansion_v2}
\end{align}
where we have defined
\beq
\label{Iketc}
\Ic_{\bk;\bq_1\cdots,\bq_n}\equiv \frac{1}{n!}\int\frac{d^3 q_1}{(2\pi)^3}\cdots \frac{d^3 q_n}{(2\pi)^3} (2\pi)^3 \delta_D\bigg(\bk-\sum_{i=1}^n \bq_i \bigg)\,.
\eeq
Notice that the functionals $ {\cal F}$, $ {\cal G}$, and $ {\cal K}$ in \re{functional} are in general non-local in space and in time. The space non-locality is due to the Poisson equation and to the fact that we consider only the divergence of the velocity fields. Moreover, the perturbative expansion of the kernels can be cast in a form that is local in time without any loss of generality, as in eqs.~\re{expansion_d2} ,\re{expansion_t2}, and \re{expansion_v2}. 
The kernels $\Fn$, $\Gn$ and $\Kn$
can be interpreted as {\it transition amplitudes} between $n$ linear  and one nonlinear fields, represented in Fig.~\ref{amp}. 
As usual, the delta function on the RHS of eq.~\eqref{Iketc} comes from assuming {\em translational invariance}  of the field equations of motion. Moreover, {\em rotational invariance} imposes that the kernels depend on  rotational invariant combinations of the momenta $\bq_i$.
In writing eqs.~(\ref{expansion_d2}--\ref{expansion_v2}) we have assumed that the only scales entering the kernels are  the external momenta $\bq_i$, i.e.~there are no other intrinsic scales in the problem (such as e.g.~the neutrino masses, massive fields in modified gravity, etc.).
Indeed, being obtained by a functional expansion around $\vphi_{\bq_i}(\eta)=0$, the kernels contain no information on the intrinsic scales of the initial power spectrum or of higher-order correlators of the initial conditions. We are also assuming no primordial non-gaussianity as, for biased tracers, it would induce a coupling between different scales which would affect the kernels (see, for instance \cite{Angulo:2015eqa}).
Moreover, the perturbative expansion is well known to break down at short scales. This is usually taken into account by suitable counterterms. We  ignore these terms for the time being and  we discuss them in sect.~\ref{sec:HD}.

The symmetry of the integration domain and of the multi-dimensional integration measure in \re{expansion_v2} translates in the requirement that the amplitude is symmetric under exchange of any pair of external momenta, i.e.,
\be
\Kn(\bq_1,\cdots, \bq_i,\cdots,\bq_j\cdots,\bq_n;\eta)= \Kn(\bq_1,\cdots, \bq_j,\cdots,\bq_i\cdots,\bq_n;\eta)\,.
\label{symb}
\ee
Next, we will consider the two  sets of symmetries to be imposed on our kernels.

\subsection{Extended Galilean Invariance}
\label{GI}

\begin{figure}[t]
\centering 
\includegraphics[width=.55\textwidth,clip]{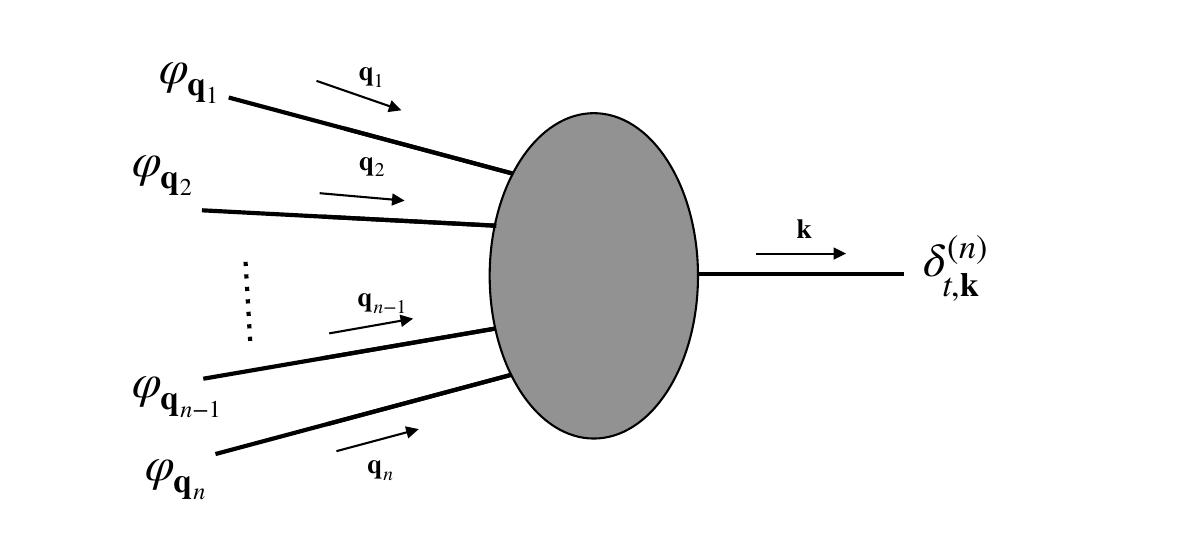}
\caption{Symbolic representation of the $\Kn(\bq_1,\cdots,\bq_n;a)$ halo amplitude.}
\label{amp}
\end{figure}
We will use the invariance under time-dependent translations to derive relations between the kernels at different orders.
In the non-relativistic limit, the dark matter fluid dynamics is invariant under  the following change of coordinates  \cite{Peloso:2013zw,Kehagias:2013yd},
\be
\label{coord_transf}
\eta \to \tilde \eta =  \eta \;, \qquad  \bx\to \tilde \bx = \bx + \bd (\eta)\,,
\ee
followed by an appropriate transformation of the density and velocity  fields,
\be
\label{deltav_transf}
 \delta (\bx, \eta) \to \tilde  \delta (\tilde \bx, \tilde \eta)  =  \delta (\bx, \eta)\qquad {\bv} (\bx, \eta)  \to \tilde {\bv}  (\tilde \bx, \tilde \eta) = {\bv} (\bx, \eta) + \HH \,\partial_\eta {\bf d}(\eta) \;, 
\ee
while the Newtonian potential transforms as 
\be
\Phi\to \Phi -    \left[ \HH \partial_\eta(\HH \partial_\eta\bd ) + \HH^2 \partial_\eta\bd  \right]  \cdot  \bx \;.
\label{pot_TT}
\ee
This is a symmetry regardless of the time dependence of $\bd$. However, to derive relations between the kernels we will impose that
 the new solution generated by the time-dependent translation is also the long-wavelength limit of a  physical mode  satisfying the equations of motion. As  shown in Appendix~\ref{A1} this assumption is not necessary but it simplifies a lot the derivation of the constraint \re{GIc}.  
 
 We can thus  consider ${\bd}(\eta)$ as  the zero momentum limit of a space-dependent field. 
  Thus, at the linear level, in Fourier space  we have the transformation
 \be
 \label{tdeq}
 \delta^{(1)}_{\bk}  \to \tilde  \delta^{(1)}_{\bk}   =  \delta_{\bk}^{(1)}   + i (2 \pi)^3 \delta_D (\bk ) \bk \cdot {\bf d}^{(1)} \;,   \qquad {\bv}_{\bk}^{(1)}   \to \tilde {\bv}_{\bk}^{(1)}   = {\bv}_{\bk}^{(1)}  + (2 \pi)^3 \delta_D (\bk ) \HH \, \partial_\eta{\bf d}^{(1)}  \;, 
\ee
where ${\bf d}^{(1)}(\eta)$ is understood as the linear component of the full displacement field ${\bf d}(\eta)$.
Using eqs.~\eqref{linfields} and \eqref{philin}, this  implies
 \be
 \label{shiftphi}
 \varphi_{\bk}  \to \tilde   \varphi_{\bk}   =   \varphi_{\bk}  + i (2 \pi)^3 \delta_D (\bk ) \bk \cdot {\bf d}^{(1)}  \;. 
\ee
The new fields  generated by the time-dependent traslation manifestly satisfy the linear continuity equation, $ \partial_\eta\tilde \delta^{(1)}_{\bk} = i \bk  \cdot \tilde{\bv}_\bk^{(1)}/\cal{H} $. 
Moreover, imposing that the new solution satisfies the Euler  equation implies \cite{Horn:2014rta}
\be
 \HH \partial_\eta(\HH \partial_\eta\bd ^{(1)}) + \HH^2 \partial_\eta\bd^{(1)}  = 4 \pi G \bar \rho a^2 \bd^{(1)} \;,
\ee
i.e., the displacement $\bd^{(1)}(\eta)$ evolves in time following the same linear growth as the original field $\vphi_\bk(\eta)$.

As shown by eq.~\eqref{pot_TT}, this  coordinate transformation permits to remove the effect of the long-wavelength gravitational potential and settle in an inertial frame, as allowed by the Equivalence Principle \cite{Creminelli:2013mca}. 
This symmetry  is not limited to the dark matter fluid dynamics. Assuming that  all species fall in the same way in the gravitational field, it can be directly applied to any biased tracer such as the galaxy distribution \cite{Creminelli:2013poa,Horn:2014rta}, replacing $\delta$ with $\delta_t$ in eq.~\re{deltav_transf},
\be
\label{deltat_transf}
 \delta_t (\bx, \eta) \to \tilde  \delta_t (\tilde \bx, \tilde \eta)  =  \delta_t (\bx, \eta) \;.
\ee
Moreover, the symmetry is non-perturbative, i.e.~it remains valid even for very short wavelengths of $\delta_t$, where  the complex baryonic  physics  is difficult to model perturbatively.

Using eq.~\re{deltat_transf}, we can write that under the coordinate transformation \re{coord_transf}, the field transforms as,
\be
 \delta_t (\bx, \eta) \to  \delta_t ( \bx - \bd,  \eta) \;,
 \ee
which,  for each Fourier mode $\bk$ can be written as
\begin{align}
\delta_{t,\bk}(\eta)& \to  e^{i \bk\cdot\bd(\eta)}\delta_{t,\bk}(\eta)  =  \sum_{m=0}^\infty  \frac{(i  \bk\cdot\bd)^m }{m!} \delta_{t,\bk}  \;. 
 \label{tGI}
 \end{align}
 Notice that both $\delta_{t,\bk}(\eta)$ and $\bd(\eta)$ in the equation above are nonlinear quantities, which we will expand perturbatively. In particular, we will first consider the effect of the linear contribution to $\bd(\eta)$, $\bd^{(1)}(\eta)$, and then the effect of the higher orders.
 \subsubsection{Leading Order}
 \label{sec:constr1}
Expanding perturbatively the left- and RHS of eq.~\re{tGI} using eq.~\eqref{expansion} and treating ${\bf d}^{(1)}(\eta)$ as the zero momentum limit of a linear field gives, for any given order $n$,
\beq
 \delta^{(n)}_{t,\bk}  \to  \sum_{m=0}^n  \frac{(i  \bk\cdot\bd^{(1)})^m }{m!} \delta^{(n-m)}_{t,\bk}   \,,
 \label{exppert}
 \eeq
 where we have omitted the dependence on $\eta$ to avoid cluttering.
This equation shows how the LHS of eq.~\eqref{expansion_v2} transforms under \re{coord_transf}. Let us now consider the RHS of eq.~\eqref{expansion_v2}. 
 If we now perform the shift \re{shiftphi} on this side of the equation we have, for the $n$-th order contribution and omitting again the $\eta$ dependence,
 \begin{align}
 & \;\Ic_{\bk;\bq_1\cdots,\bq_n}\, \Kn(\bq_1,\cdots,\bq_n) \vphi_{\bq_1} \cdots  \vphi_{\bq_n} \nonumber\\
 & \to  \Ic_{\bk;\bq_1\cdots,\bq_n}\, \Kn(\bq_1,\cdots,\bq_n)\, \sum_{m=0}^{n}  \left(\begin{array}{c} n\\m\end{array}\right)  \, \epsilon_{\bq_1} \cdots \epsilon_{\bq_m} \vphi_{\bq_{m+1}} \cdots  \vphi_{\bq_n} \,,
 \label{shpt}
 \end{align}
 where for compactness we have defined $\epsilon_\bk(\eta) \equiv i (2 \pi)^3 \delta_D(\bk) \bk \cdot \bd^{(1)}(\eta)$. Notice that the $m=0$ term in the second line coincides with $ \delta^{(n)}_{t,\bk}(\eta)$.

 We can now equate the  ${\cal O}(({\bf d}^{(1)})^m)$ contribution on the RHS of eq.~\re{exppert}, i.e., 
 \begin{align}
 & \frac{i^m}{m!}\left(\bk\cdot\bd^{(1)} \right)^m \, \delta^{(n-m)}_{t,\bk} = \frac{i^m}{m!}\left(\bk\cdot\bd^{(1)} \right)^m\, \Ic_{\bk;\bfp_1\cdots,\bfp_{n-m}}\, \Knmenom (\bfp_1,\cdots,\bfp_{n-m})\, \vphi_{\bfp_1} \cdots  \vphi_{\bfp_{n-m}} \,,
 \end{align}
 with that on  the RHS of eq.~\re{shpt}, and obtain a relation between the $n$-th order kernel and the $(n-m)$-th one,
 \begin{align}
 &  \int d^3q_1\cdots d^3q_m (2\pi)^3\delta_D\left(\bk-{\bf Q}_{n,0}\right) \, \Kn(\bq_1,\cdots,\bq_m,\bq_{m+1}\cdots\bq_n ) \bq_1\cdot {\bf d}\,\delta_D(\bq_1) \cdots \bq_m\cdot {\bf d} \,\delta_D(\bq_m)\nonumber\\
 & =(2\pi)^3 \delta_D\left(\bk-{\bf Q}_{n,m}\right)  \Knmenom (\bq_{m+1},\cdots,\bq_{n}) \,\left(\bk\cdot\bd \right)^m\,,
 \label{ktkt}
 \end{align}
 where  we have defined 
 \be
 \label{QQ}
 {\bf Q}_{n,m} \equiv \sum_{i=m+1}^n \bq_i \;.
 \ee
Due to the momentum delta functions in the first line and the arbitrariness of the shift $\bd(\eta)$, the above relation fixes the soft limit of the $n$-th order kernel as
 \begin{align}
 \lim_{\bq_1,\cdots, \bq_m \to 0} &\;q_1^{i_1}\cdots q_m^{i_m}\, \Kn(\bq_1,\cdots,\bq_m,\bq_{m+1}\cdots\bq_n) =   Q_{n,m}^{i_1}\cdots Q_{n,m}^{i_m} \, \Knmenom (\bq_{m+1}\cdots\bq_n) +{\cal O}(q)\,.
 \end{align}
 By further contracting by $q_1^{i_1}\cdots q_m^{i_m}$ we get a series of constraints on the pole structure of the kernel, which reads
 \begin{align}
  \lim_{\bq_1,\cdots, \bq_m \to 0} &\Kn(\bq_1,\cdots,\bq_m,\bq_{m+1}\cdots\bq_n ) \nonumber\\
  &= \frac{\bq_1\cdot {\bf Q}_{n,m}}{q_1^2}\cdots  \frac{\bq_m\cdot {\bf Q}_{n,m}}{q_m^2}\, \Knmenom (\bq_{m+1}\cdots\bq_n) +{\cal O}((1/q)^{m-1})\,.
  \label{GIc}
 \end{align}
 These constraints enforce the symmetries of the equations of motions, namely EGI. They are closely related, but not completely equivalent, to the consistency relations of the LSS  \cite{Peloso:2013zw,Kehagias:2013yd}. Indeed, the latter  involve fully non-perturbative correlators, as opposed to the intrinsically perturbative kernels appearing in \re{GIc}.  On the other hand, in order to derive the consistency relations, additional assumptions on the state of the system are needed, namely adiabaticity and gaussianity of the initial conditions. These are not necessary here, as least for unbiased tracers, as we implement the symmetries of the equations of motion, regardless of the state.

  \subsubsection{Next to Leading Order}
  \label{NLOsec}
Eq.~\re{tGI} can also be read as follows. If the initial conditions, once evolved, produce a long mode $\bd(\eta)$, then, in the limit of infinitely long wavelength,  its effect on the nonlinear tracer field should factorise as 
 \be 
 e^{i \bk\cdot\bd(\eta)}\delta_{t,\bk}(\eta)\,,
 \label{facto}
 \ee
 where $\delta_{t,\bk}(\eta)$ is  $\bd(\eta)$-independent. This statement holds beyond the assumption we made in the previous subsection, namely, that $\bd(\eta)$ is the long wavelength limit of a linear field. Indeed, the coupling of two linear modes, $\varphi_{\bq_1}$ and  $  \varphi_{\bq_2}$ gives a displacement field, in Fourier space,
 \begin{align}
 \tilde \bd^{(2)}_\bq(\eta)& =\int^{\tau(\eta)} d\tau' \,\bf{v}_\bq^{(2)}(\tau')\nonumber\\ 
  &=- i \frac{\bq}{q^2}  \int^\eta d \eta' f_+(\eta') \Ic_{\bq;\bq_1,\bq_2}\, \Gtwo(\bq_1,\bq_2;\eta') \vphi_{\bq_1}(\eta')  \vphi_{\bq_2}(\eta')\,,\nonumber\\
  &=-   i \frac{\bq}{q^2}\int^\eta d \eta' f_+(\eta')  \frac{D_+(\eta')^2}{D_+(\eta)^2}\Ic_{\bq;\bq_1,\bq_2}\, \Gtwo(\bq_1,\bq_2;\eta') \vphi_{\bq_1}(\eta)  \vphi_{\bq_2}(\eta)\,.
  \label{dt}
 \end{align}
 
In the $\bq=\bq_1+\bq_2 \to 0$ limit, the coupling between these two modes gives a contribution to the nonlinear field of the form \re{facto}. The lowest contribution is of  third order,
\beq
i \bk\cdot\bd^{(2)}(\eta) \delta_{t,\bk}^{(1)}(\eta)\,,
\label{d2}
\eeq
where the `zero mode' displacement is defined as
\be
\bd^{(2)}(\eta) \equiv \frac{1}{V}\int_V d^3 x \int \frac{d^3 q}{(2\pi)^3} e^{-i \bq\cdot\bx} \,\tilde \bd^{(2)}_\bq(\eta),
\ee
with $V=(2\pi)^3 \delta_D(k=0)$.
Using \re{dt} in \re{d2} then gives
\begin{align}
&i \bk\cdot\bd^{(2)}(\eta) \delta_{t,\bk}^{(1)}(\eta) \nonumber\\
&\quad= \frac{1}{V}{\cal I}_{0;\bq_1,\bq_2}\frac{\bk\cdot \bq_{12}}{q^2_{12}}\int^\eta d \eta' f_+(\eta')  \frac{D_+(\eta')^2}{D_+(\eta)^2}\, \Gtwo(\bq_1,\bq_2;\eta') \vphi_{\bq_1}(\eta)  \vphi_{\bq_2}(\eta) \Kone(\bk;\eta)\vphi_{\bk}(\eta)\,,
\label{id2}
\end{align}
where we have defined $\bq_{ij}\equiv \bq_i+\bq_j$.
NLO EGI requires that this contribution is contained in the expression for $\delta_{t,\bk}^{(3)}(\eta)$ from eq.~\re{expansion_v2}. In order to isolate it, we insert the  identity
\begin{align}
&1=\frac{(2\pi)^3}{V}\left( \delta_D(\bq_{12})+  \delta_D(\bq_{23})+  \delta_D(\bq_{31})\right) +\left[1-\frac{(2\pi)^3}{V}\left( \delta_D(\bq_{12})+  \delta_D(\bq_{23})+  \delta_D(\bq_{31})\right)  \right]
\end{align}
in the momentum integral. From the first term of the identity we get 
\begin{align}
&\delta^{(3)}_{t,\bk}(\eta) \supset \Ic_{\bk;\bq_1,\bq_2,\bq_3}\, \Kthree(\bq_1,\bq_2,\bq_3;\eta) \vphi_{\bq_1}(\eta)\vphi_{\bq_2}(\eta)\vphi_{\bq_3}(\eta)\frac{(2\pi)^3}{V}\left( \delta_D(\bq_{12})+  \delta_D(\bq_{23})+  \delta_D(\bq_{31})\right)\nonumber\\
&\qquad\qquad= \frac{1}{V}\; \Ic_{0;\bq_1,\bq_2}\, \Kthree(\bq_1,\bq_2,\bk;\eta)  \vphi_{\bq_1}(\eta)\vphi_{\bq_2}(\eta)\vphi_{\bk}(\eta),
\end{align}
which should be identified with \re{id2}, therefore leading to the relation
\begin{align}
&\lim_{\bq_{12}\to 0 }\Kthree(\bq_1,\bq_2,\bk;\eta) \; \supset \; \frac{\bk\cdot \bq_{12}}{q^2_{12}}\Kone(\bk;\eta) \int^\eta d \eta' f_+(\eta')  \frac{D_+(\eta')^2}{D_+(\eta)^2} \Gtwo(\bq_1,\bq_2;\eta') \,.
\label{nlo}
\end{align}
Notice that, as we will discuss in Sect.~\ref{MMC}, mass and momentum conservation implies $G_2(\bq_1,\bq_2;\eta) \sim q_{12}^2$ as $q_{12}\to 0$. Therefore, unlike eq.~\re{GIc}, the kernels have no pole when the sum of two (or more, see next subsection) momenta vanish. Moreover, the term at the RHS of eq.~\re{nlo} might also not  be the leading one in this limit, as, for a generic tracer, constant contributions are also present. Nevertheless this equation fixes the structure of the terms containing the $\bk\cdot\bq_{12}/q_{12}^2$ combination in the $q_{12}\to 0$ limit which, as will see, provides meaningful constraints to the kernels.

 \subsubsection{N$^{l-1}$LO}
It is possible to generalize eq.~\re{nlo} to a general order of the kernel and with the sum of an arbitrary number of momenta going to zero. Exploiting the coupling between $l$ linear modes, one can expand the displacement field as
\beq
\label{gendisp}
\tilde{\bd}_\bq^{(l)}(\eta) = -i\frac{\bq}{q^2}\int^{\eta}d\eta' f_{+}(\eta')\left(\frac{D_+(\eta')}{D_+(\eta)}\right)^{l}{\cal I}_{\bq;\bq_1,\dots,\bq_l} \Gl (\bq_1,\dots,\bq_l;\eta')\vphi_{\bq_1}(\eta)\dots\vphi_{\bq_l}(\eta),
\eeq
from which we can obtain the general relation for the $n$-th order kernel
\begin{align}
\label{genGIshort}
\lim\limits_{{\bf Q}_{l,0}\to0} &\Kn (\bq_1,\dots,\bq_l,\bq_{l+1},\dots,\bq_n) \supset \nonumber\\
&\frac{\bk\cdot{\bf Q}_{l,0}}{ Q_{l,0}^2}\int^{\eta}d\eta'\,f_+(\eta')\left(\frac{D_+(\eta')}{D_+(\eta)}\right)^l  \Gl (\bq_{1},\dots,\bq_l;\eta')\Knmenol (\bq_{l+1},\dots,\bq_n;\eta)\,,
\end{align}
where we recall that ${\bf Q}_{l,0} = \sum_{i = 1}^l \bq_i$ (see eq.~\re{QQ}).
The number of momenta $l$, for which the sum is set to zero in eq.~\re{genGIshort}, gives the order of the perturbative expansion for the long-wavelength displacement. Considering, for example, $l=1$ and iterating $m$ times eq.~\re{genGIshort}, gives the leading order extended Galilean invariance expressed in eq.~\re{GIc}. To obtain the NLO order one should choose $l=2$, meaning that the sum of two internal momenta is set to zero, so that the first correction of this type arise for kernels $\Kn$ with $n\geq3$, leading to eq.~\re{nlo}. The explicit  calculation of the NNLO condition, which starts to play a role at the fourth PT order, is presented in Appendix ~\ref{nnlosec}.

 \subsection{Mass and momentum conservation}
 \label{MMC}

 In  Sec.~\ref{GI} we have discussed the behavior of the kernels $\Kn (\bq_1,\cdots\bq_n;\eta) $ as one or more of the  momenta of the linear fields, $q_i$, vanishes. In this section we inspect the opposite limit, in which the momentum of the nonlinear field, $k$, is $\ll q_i$ for all $i$'s.
We consider a tracer that satisfies mass and momentum conservation, such as for instance the dark matter density contrast.

Mass conservation imposes that,
\beq
\int d^3 x \,  \delta(\bx,\eta)=0\,,
  \label{delta0}
\eeq
while momentum conservation, namely that the center of mass of the dark matter distribution is fixed, imposes that, 
 \be
 \int d^3 x x^i\,  \delta(\bx,\eta)=0\,.
 \label{deltap}
 \ee
Inserting \re{expansion} in these two conditions and imposing that they are true independently of the linear initial linear fields give two independent constraints on the kernels,
 \begin{align}
 &\lim_{\bfQ_{n,0}\to 0 } \Fn(\bq_1,\cdots\bq_n;\eta) =0\,,
 \nonumber\\
 &\lim_{\bfQ_{n,0}\to 0}  \frac{\partial }{\partial q_1^i} \Fn(\bq_1,\cdots\bq_n;\eta)  =0\,,
 \label{km}
 \end{align}
 where the limit is taken by keeping all the individual $q_i$'s non vanishing.
The conditions above ensure that the density contrast decouples as 
 \be
 \delta_{\bk}(\eta)={\cal O}\left({k^2}/{q_i^2}\right)\,,
 \label{deco}
 \ee
 when the external momentum is much smaller than the $q_i$'s,  i.e.~$k\ll q_i$, as implied by general arguments on momentum conservation \cite{Zeldo,PeebBook,Mercolli:2013bsa,Abolhasani:2015mra}.

The two conditions eq.~\re{delta0} and \re{deltap} hold for the matter density contrast and velocity divergence \cite{Mercolli:2013bsa}, so that eq.~\eqref{km} apply to the matter and velocity kernels , i.e.~$\Fn$ and $\Gn$. However, they do not hold for a generic tracer, such as the galaxy number density, which does not satisfy a conservation equation and for which eqs.~\eqref{km} do not apply. Therefore, in the following we will denote as `matter kernels' the ones satisfying the conditions in eq.~\re{km}. We turn now to derive their explicit forms by imposing the symmetries discussed above.

 \section{Matter Kernels}
  \label{matterk}

 \subsection{Imposing the constraints}    

    
Here we will  implement all the constraints discussed in Sec.~\ref{constr}, including the last one,  and derive the expressions of the matter kernels, i.e.~for $\Fn$ and $\Gn$. The details of these calculations are explicitly given in App.~\ref{app:kernels}.
We will use isotropy
 to write  the kernels in terms of rotationally invariant objects constructed from the external momenta. 
For $n=1$ we have one invariant, i.e.~$q_1^2=k^2$. For $n\ge 2$ we need $3 n -3 $ of them and we can take them from the $n(n+1)/2$ scalar products $\bq_l \cdot \bq_m$. For $n=2$ these are $q_1^2$, $q_2^2$, $\bq_1 \cdot \bq_2$, and for $n=3$ they are $q_1^2$, $q_2^2$, $q_3^2$, $\bq_1 \cdot \bq_2$, $\bq_1 \cdot \bq_3$ and $\bq_2 \cdot \bq_3$.  
 We will then construct, up to third order, the most general dimensionless functions built out of the $\bq_i$ momenta satisfying the above conditions.
 We  restrict to rational functions, which is consistent with the perturbative nature of the present analysis, as expressed by eq.~\re{expansion}.

  For $n=1$, the kernel depends  only on one momentum, $\bq$, but  isotropy  implies that the dimensionless rotational invariant is just a numerical constant, implying
 \be
 \label{F1}
 \Fone(\bq_1;\eta) = 1\,.
 \ee
 
  For $n=2$ there are two independent external momenta. 
 The requirement discussed in sect. \ref{GI} above restricts the order of the poles, when one of the momenta vanishes, at most to the first. Therefore, the rotational invariants we can build out of $\bq_1$ and $\bq_2$ are limited to four:
 \be
 1\,,\qquad \frac{\bq_1\cdot\bq_2}{q_1^2}  , \qquad  \frac{\bq_1\cdot\bq_2}{q_2^2}  , \qquad \frac{(\bq_1\cdot\bq_2)^2}{q_1^2 q_2^2} \;.
 \ee
We find it convenient, instead, to use as basis functions the following combinations
\be
\label{bas}
\begin{split}
1, \qquad \gamma (\bq,\bfp)  = 1 - \frac{(\bq \cdot \bfp)^2}{ q^2 p^2} \;, \qquad \beta(\bq,\bfp)\equiv\frac{|\bq+\bfp|^2\,\bq\cdot\bfp}{2 q^2p^2} \;, \qquad  \alpha_a(\bq,\bfp) = \frac{\bq \cdot \bfp}{ q^2}- \frac{\bfp \cdot \bq}{ p^2}\,,
\end{split}
\ee
where the last combination appears only at $n\ge 3$.
Indeed, requiring  total symmetrization, eq.~\re{symb}, restricts the invariants to three, leading to the most generic form for the  second-order kernel,
 \begin{align}
  \Ftwo(\bq_1,\bq_2;\eta) = \,\AC_{0}^{(2)}(\eta)+\AC_{1}^{(2)}(\eta) \,\gamma(\bq_1,\bq_2) +\AC_{2}^{(2)}(\eta) \,\beta(\bq_1,\bq_2) \,.
  \label{kappa2}
 \end{align}
 Imposing the constraints  discussed in sect.~\ref{constr} gives $\AC_{ 0}^{(2)}=0$ from mass conservation, and $\AC_{2}^{(2)}=2$ from  EGI (see App.~\ref{app:kernels}), so that
the most general matter kernel at $n=2$ is given by
 \beq
   \Ftwo(\bq_1,\bq_2;\eta) =2\beta(\bq_1,\bq_2) + \AC_{1}^{(2)}(\eta) \, \gamma(\bq_1,\bq_2)  \,,
   \label{kernel2}
 \eeq
 which contains only one independent function of time,  $\AC_{1}^{(2)}(\eta)$.

 For $n=3$,  the most general matter kernel can be obtained by  taking the direct product between the independent structures built out of $\bq_1$ and $\bq_2$ and those built out of $\bq_{12}$ and $\bq_3$, and then symmetrizing over $\bq_1$, $\bq_2$, and $\bq_3$,
 \beq
 \left(1, \gamma(\bq_1,\bq_2), \beta(\bq_1,\bq_2)\right) \otimes\left(1,\gamma(\bq_{12},\bq_3),\alpha_a(\bq_{12},\bq_3),\beta(\bq_{12},\bq_3) \right)+  {\mathrm{ cyclic} }\,.
 \label{dirprod}
 \eeq
The $\alpha_a(\bq_1,\bq_2)$ term in the first factor is absent due to symmetrization. The symmetrization process is completed by adding the remaining two cyclic permutations. 
Thus, the most general structure of the third-order kernel starts with 12 terms. Suppressing the dependence on $\eta$ to avoid clutter, this reads
\begin{align}
\Fthree(\bq_1,\bq_2,\bq_3) =&\,\,\frac{1}{3} \AC_{ 0}^{(3)} + \AC_{ 1}^{(3)} \gamma(\bq_1,\bq_2) +  \AC_{ 2}^{(3)}  \gamma(\bq_{12},\bq_3) + \AC_{ 3}^{(3)} \beta(\bq_1,\bq_2) +  \AC_{ 4}^{(3)} \beta(\bq_{12},\bq_3) \nonumber\\
&+ \AC_{ 5}^{(3)}\gamma(\bq_1,\bq_2)  \gamma(\bq_{12},\bq_3)+  \AC_{ 6}^{(3)} \beta(\bq_1,\bq_2)  \beta(\bq_{12},\bq_3) + \AC_{ 7}^{(3)} \gamma(\bq_1,\bq_2)  \beta(\bq_{12},\bq_3)\nonumber\\
&+  \AC_{ 8}^{(3)} \beta(\bq_1,\bq_2)  \gamma(\bq_{12},\bq_3) + \Big( \AC_{ 9}^{(3)}+\AC_{ 10}^{(3)} \gamma(\bq_1,\bq_2) +\AC_{ 11}^{(3)} \beta(\bq_1,\bq_2)  \Big)\,\alpha_a(\bq_{12},\bq_3) \nonumber\\
&+ {\mathrm{ cyclic}}\,.
\label{k3g}
\end{align}

Imposing the symmetries discussed in sect.~\ref{constr}   provides  10 independent conditions on the 12 coefficients appearing in \re{k3g} (see again  App.~\ref{app:kernels} for details), and thus restricts the total number of independent coefficients to three, i.e.~
\be
\left\{\AC_{ 1}^{(2)} ,\, \AC_{ 5}^{(3)} ,\; \AC_{ 10}^{(3)} \right\}\,.
\label{matterCC}
\ee
In this case the kernel reads
\begin{align}
 \Fone(\bq_1) &= 1\, \\
 \label{matterK2} \Ftwo(\bq_1,\bq_2) &=2\beta(\bq_1,\bq_2) + \AC_{1}^{(2)} \, \gamma(\bq_1,\bq_2)  \,, \\
\Fthree(\bq_1,\bq_2, \bq_3) &= 2\,\beta(\bq_1,\bq_2)\beta(\bq_{12},\bq_3) + \AC_{5}^{(3)}   \gamma(\bq_1,\bq_2)  \gamma(\bq_{12},\bq_3)  \nonumber \\
& - 2 \,\left(\AC_{10}^{(3)}   -  \hh \right)\gamma(\bq_1,\bq_2)\beta(\bq_{12},\bq_3) + 2 (\AC_{1}^{(2)}  + 2\,\AC_{10}^{(3)} -  \hh) \beta(\bq_1,\bq_2)\gamma(\bq_{12},\bq_3) \nonumber \\
& + \AC_{10}^{(3)}\gamma(\bq_1,\bq_2)\alpha_a(\bq_{12},\bq_3)+ {\mathrm{ cyclic}}\,,
\label{matterK}
\end{align}
where the time-dependent coefficient $h$ is not independent and is defined below (eq.~\eqref{hhdef}).

All the kernel relations above hold analogously for the velocity divergence: $\Gn$ can be rewritten up to the third order as eq.~\re{F1}, \re{kernel2} and \re{matterK}, i.e.,
\begin{align}
 \Gone(\bq_1) &= 1\, \\
\Gtwo(\bq_1,\bq_2) &=2\beta(\bq_1,\bq_2) + \BC_{1}^{(2)} \, \gamma(\bq_1,\bq_2)  \,, \label{G2matter}\\
\Gthree(\bq_1,\bq_2, \bq_3) &= 2\,\beta(\bq_1,\bq_2)\beta(\bq_{12},\bq_3) + \BC_{5}^{(3)}   \gamma(\bq_1,\bq_2)  \gamma(\bq_{12},\bq_3)  \nonumber  \\
&  - 2 \, \left(\BC_{10}^{(3)} -  \hh  \right)\gamma(\bq_1,\bq_2)\beta(\bq_{12},\bq_3)   + 2 (\BC_{1}^{(2)}  + 2\,\BC_{10}^{(3)} -  \hh ) \beta(\bq_1,\bq_2)\gamma(\bq_{12},\bq_3) \nonumber \\
& + \BC_{10}^{(3)}\gamma(\bq_1,\bq_2)\alpha_a(\bq_{12},\bq_3)+ {\mathrm{ cyclic}}\,.
\label{matterKG}
\end{align}
The coefficient $\hh$ is defined  as
\be
\hh(\eta) \equiv   \int^\eta d\eta'\, f_+(\eta')\left[\frac{D_+(\eta')}{D_+(\eta)}\right]^2 \BC_{1}^{(2)}(\eta') \;,
\label{hhdef}
\ee
with $\BC_{1}^{(2)}$ being the time-dependent coefficient of $\Gtwo$  in eq.~\eqref{G2matter}.

\subsection{Time dependence}
\label{time-dep}

 The discussion so far did not assume any specific cosmological model, namely a background evolution and a definite form of the system of Euler, continuity and Poisson equations. In this section we show how fixing a cosmology allows to derive the time dependence on the three parameters  in eq.~\re{matterCC} and the corresponding ones for the velocity field.  We will do this  first in the standard  $\Lambda$CDM case and then for a particular modified gravity model.

\subsubsection{$\Lambda$CDM}
\label{LCDM}

Extending the notation of eq.~\re{linfields}  to nonlinear fields, and switching the time variable from $\eta$ to 
\beq
\chi\equiv \log\frac{D_+(\eta)}{D_+(0)}\,,
\eeq
we can write the equation of motion for the $n$-th order ($n>1$) one as (see e.g.~\cite{Crocce:2005xy}), 
\be
\left(\delta_{\lambda\lambda'} \partial_\chi + \Omega_{\lambda\lambda'}(\chi) \right)\Psi^{\lambda',(n)}_\bk(\chi) = {\cal I}_{\bk;\bq_1,\bq_2} \gamma_{\lambda\lambda'\lambda''}(\bq_1,\bq_2) \sum_{m=1}^{n-1} \Psi^{\lambda',(m)}_{\bq_1}(\chi)  \Psi^{\lambda'',(n-m)}_{\bq_2}(\chi)\,,
\label{eomL}
\ee
where $\Psi^\lambda$ ($\lambda = 1,2$) is the doublet
\be 
\Psi^1(\bx,\eta ) \equiv \delta (\bx, \eta)\,,\qquad \Psi^2(\bx,\eta )\equiv  \Theta (\bx,\eta) \,.
\ee
Moreover, assuming $\Lambda$CDM, we have
\be 
{\bf \Omega}(\chi)=\left(\begin{array}{cc} 0&-1\\-\frac{3}{2}\frac{\Omega_m(\chi)}{f_+^2(\chi)}& \frac{3}{2}\frac{\Omega_m(\chi)}{f_+^2(\chi)}-1\end{array}\right)\,,
\ee
\be
\begin{split}
 \gamma_{121}(\bq_1,\bq_2)&= \gamma_{112}(\bq_2,\bq_1)= \frac12 \left( 1 + \frac{\bq_1 \cdot \bq_2}{q_1^2} \right) \,, \\
 \gamma_{222}(\bq_1,\bq_2)&=\beta(\bq_1,\bq_2)\,,
\end{split}
\ee
and zero otherwise.

Using eq.~\re{expansion} we obtain the evolution equations of the kernels. 
At order  $n=2$, noting that $\gamma_{112}(\bq_1,\bq_2) + \gamma_{121}(\bq_1,\bq_2) = \gamma (\bq_1,\bq_2) + \beta (\bq_1,\bq_2)$ we get
\be
\begin{split}
&\left(\partial_\chi+2\right)\Ftwo(\bq_1,\bq_2;\chi) -  \Gtwo(\bq_1,\bq_2;\chi)=2 \big[ \gamma(\bq_1,\bq_2) + \beta(\bq_1,\bq_2) \big]\,,\\
&\left(\partial_\chi +1+ \frac{3}{2}\frac{\Omega_m(\chi)}{f_+^2(\chi)} \right)\Gtwo(\bq_1,\bq_2;\chi) - \frac{3}{2}\frac{\Omega_m(\chi)}{f_+^2(\chi)}  \Ftwo(\bq_1,\bq_2;\chi)=2\,\beta(\bq_1,\bq_2)\,.
\end{split}
\ee
 Moreover, using the form of the kernels in eq.~\re{kernel2}  and selecting the growing mode initial condition, we get the system of two coupled equations for the kernel  coefficients,
\be
\label{eomLb}
\begin{split}
&(\partial_\chi +1) \AC_{1}^{(2)} =2 -  \AC_{1}^{(2)}+\BC_{1}^{(2)}\,,\\
&(\partial_\chi+1) \BC_{1}^{(2)}  = \frac{3}{2}\frac{\Omega_m }{f_+^2} \left( \AC_{1}^{(2)}- \BC_{1}^{(2)}\right)\,.
\end{split}
\ee
These equations are to be solved with the  initial conditions  
\be
\label{icond}
 \lim_{\chi\to-\infty} \{\AC_{1}^{(2)},\, \BC_{1}^{(2)} \} =\{\AC_{1,{\rm EdS}}^{(2)},\, \BC_{1, {\rm EdS}}^{(2)} \}\,,
 \ee 
 where the Einstein de Sitter values are the solutions corresponding to $\Omega_m/f_+^2=1$ which are finite in the $\chi\to-\infty$ limit, 
 \beq
\AC_{1,{\rm EdS}}^{(2)} = \frac{10}{7}\,,\qquad \BC_{1,{\rm EdS}}^{(2)} = \frac{6}{7}\,.
\label{EdS2}
\eeq 
The  analytic solutions are given by 
\be
\begin{split}
\AC_{1}^{(2)}(\chi)&=2-e^{-\chi}\int_{-\infty}^\chi d\chi'\,e^{\chi'} \Delta \AC_1^{(2)}(\chi) \,,\\
\BC_{1}^{(2)}(\chi)&= \AC_{1}^{(2)}(\chi) - \Delta \AC_1^{(2)}(\chi)\,,
\label{C21}
\end{split}
\ee
where
\be
\Delta \AC_1^{(2)}(\chi)= 2 \,e^{-\chi}\int_{-\infty}^\chi d\chi'\; e^{\chi' - \int_{\chi'}^\chi d\chi''\;\left(1+\frac{3}{2}\frac{\Omega_m(\chi'')}{f_+^2(\chi'')} \right)}\,.
\label{C22}
\ee
 \begin{figure}[t]
\centering 
\includegraphics[width=.65\textwidth,clip]{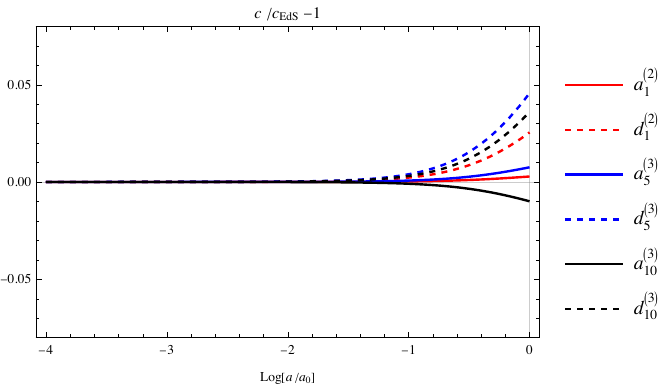}
\caption{Evolution of the coefficients of the kernels in $\Lambda$CDM (for $\Omega_m^0=0.27$),  normalized to the EdS limit.}
\label{coeffs}
\end{figure}

Going to the next order, at $n=3$, and equating the coefficients of the different structures in $\gamma$, $\beta$ and $\alpha_a$, we get the coupled evolution equations for the 6  independent parameters appearing in the kernels for the density and velocity fields.
These, as well as their analytic solutions, are written in App.~\ref{ansols}. An example of their behavior is given in Fig.~\ref{coeffs}. Notice that the deviations of the velocity coefficients (dashed lines) from the EdS is  of order  4\% at low redshifts, and is more pronounced than that of matter (solid lines).

Again,  setting  $\Omega_m/f_+^2=1$, the analytical solutions  recover the results of the EdS limit for the third-order density-field kernels,
\be 
\AC_{5,\rm{EdS}}^{(3)} =\frac{8}{9} \,, \qquad
 \AC_{10,\rm{EdS}}^{(3)} =-\frac{1}{9}\,,
 \label{EdSd}
\ee
and the third-order velocity ones,
 \be
 \BC_{5,\rm{EdS}}^{(3)} =\frac{8}{21}\,, \qquad
 \BC_{10,\rm{EdS}}^{(3)} =-\frac{1}{21}\,.
  \label{EdSv}
 \ee
\subsubsection{nDGP}
\label{bLCDM}
In theories beyond $\Lambda$CDM, where the linear growth is still scale-independent, like for instance in the normal branch of the Dvali, Gabadadze and Porrati (nDGP) braneworld model \cite{Dvali:2000xg} or in the EFT of dark energy beyond linear order \cite{Cusin:2017mzw,Cusin:2017wjg} (see  \cite{Bose:2018orj} for a comparison with simulations), new nonlinear couplings emerge. Their effect on the time-dependence of the coefficients can be derived by including them on the RHS of eq.~\re{eomL}, at the appropriate order.

 In particular, in the  nDGP model  the Poisson equation in Fourier space is modified as  \cite{Koyama:2009me}
\be
\label{modpois}
-\frac{k^2}{{\cal H}^2}\Phi(\bk,\eta) = \frac{3}{2}\Omega_m(\eta) \mu(\eta) \delta_\bk(\eta) + S(\bk,\eta),
\ee
where the function $\mu$ represents the linear modification to the Poisson equation and it is given by 
\be
\label{mu}
\mu(\eta) \equiv 1 + \frac{1}{3{\cal B}(\eta)}, \qquad {\cal B}(\eta) \equiv 1 + \frac{1}{3\sqrt{\Omega_{rc}}}\frac{{\cal H}}{{\cal H}_0}e^{-\eta}\left(2 + \frac{d\log{\cal H}}{d\eta}\right). 
\ee
The new parameter entering in eq.~\re{mu} is given by $\Omega_{rc} \equiv 1/\left(4 r_c^2 H_0^2\right)$, where $r_c$ is the crossover scale, which parametrizes the ratio between the five-dimensional Newton's gravitational constant and the four-dimensional one.

In eq.~\re{modpois} the non-linear source $S(\bk,\eta)$ expanded to third order in perturbation theory is given by \cite{Bose:2016qun}
\begin{align}
\label{source}
S(\bk,\eta) &= 2 \mu_2(\eta) \left(\frac{3}{2}\Omega_m (\eta) \right)^2  \Ic_{\bk;\bq_1 \bq_2} \gamma(\bq_1,\bq_2)\delta_{\bq_1} (\eta)  \delta_{\bq_2} (\eta) \\
&+6 \mu_{3}(\eta) \left(\frac{3}{2}\Omega_m(\eta) \right)^3  \Ic_{\bk;\bq_1 \bq_2 \bq_3} 
\gamma(\bq_2,\bq_3 )\gamma(\bq_1,\bq_{23})\delta_{\bq_1}(\eta)  \delta_{\bq_2}(\eta)  \delta_{\bq_3}(\eta)  \,,\nonumber
\end{align}
where $\gamma(\bq_1,\bq_2)$ is defined in eq.~\re{bas} and 
\be
\label{munl}
\mu_2(\eta) \equiv -\frac{{\cal H}^2(\eta) }{{\cal H}_0^2}e^{-2\eta}\frac{1}{2\Omega_{rc}}\left(\frac{1}{3{\cal B} (\eta)}\right)^3\,,\qquad \mu_3(\eta) \equiv \frac{{\cal H}^4(\eta)}{{\cal H}_0^4}e^{-4\eta}\frac{1}{2\Omega_{rc}}\left(\frac{1}{3{\cal B}(\eta)}\right)^5\,.
\ee
General relativity is recovered when $\mu = 1$ and $\mu_2 = \mu_3 = 0$.

With these specifications, the analog for nDGP of 
the equations of motions eq.~\re{eomL} for $\Lambda$CDM is
\begin{align}
\label{eomnDGP}
\left(\delta_{\lambda\lambda'} \partial_\chi + \Omega_{\lambda\lambda'} \right)\delta^{\lambda',(n)}_\bk = &{\cal I}_{\bk;\bq_1,\bq_2} \gamma_{\lambda\lambda'\lambda''}(\bq_1,\bq_2) \sum_{m=1}^{n-1} \delta^{\lambda',(m)}_{\bq_1}  \delta^{\lambda'',(n-m)}_{\bq_2}\\
&+\frac{\mu_2}{f_+^2}\left(\frac{3}{2}\Omega_m\right)^2{\cal I}_{\bk;\bq_1,\bq_2} \gamma(\bq_1,\bq_2) \sum_{m=1}^{n-1} \delta^{(m)}_{\bq_1}(\chi)  \delta^{(n-m)}_{\bq_2}(\chi)\nonumber\\
&+\frac{\mu_3}{f_+^2}\left(\frac{3}{2}\Omega_m\right)^3{\cal I}_{\bk;\bq_1,\bq_2,\bq_3} \gamma(\bq_1,\bq_2) \sum_{m=1}^{n-2} \sum_{l=1}^{n-2} \delta^{(m)}_{\bq_1}  \delta^{(l)}_{\bq_2}\delta^{(n-m-l)}_{\bq_3}\nonumber\,,
\end{align}
where the explicit dependence on $\chi$ has been removed to reduce clutter and 
\be 
{\bf \Omega}=\left(\begin{array}{cc} 0&-1\\-\frac{3}{2}\frac{\Omega_m }{f_+^2 }\mu& \frac{3}{2}\frac{\Omega_m}{f_+^2}\mu-1\end{array}\right)\,.
\ee
Again, using eq.~\re{expansion} and the one for the velocity field, we obtain the evolution equations for the kernels in this modified gravity scenario. At order $n=2$ and using eq.~\re{kernel2} for the kernels we obtain 
\be
\label{eomnDGPb}
\begin{split}
&(\partial_\chi +1) \AC_{1}^{(2)} =2 -  \AC_{1}^{(2)}+\BC_{1}^{(2)}  + 2 \frac{\mu_2}{f_+^2}\left(\frac{3}{2}\Omega_m\right)^2\,,\\
&(\partial_\chi+1) \BC_{1}^{(2)}  = \frac{3}{2}\frac{\Omega_m }{f_+^2} \left( \AC_{1}^{(2)}- \BC_{1}^{(2)}\right) + 2 \frac{\mu_2}{f_+^2}\left(\frac{3}{2}\Omega_m\right)^2\,,
\end{split}
\ee
to be solved with the same initial condition expressed in eq.~\re{icond}. The analytic solution is
\be
\begin{split}
\AC_{1}^{(2)}(\chi)&=2-e^{-\chi}\int_{-\infty}^\chi d\chi'\,e^{\chi'}  \left[ \Delta \AC_1^{(2)}(\chi')  - 2\frac{\mu_2}{f_+^2}\left(\frac{3}{2}\Omega_m\right)^2\right]\,,\\
\BC_{1}^{(2)}(\chi)&= \AC_{1}^{(2)}(\chi) - \Delta \AC_1^{(2)}(\chi)\,,
\label{C21nDGP}
\end{split}
\ee
where in nDGP 
\be
\Delta \AC_1^{(2)}(\chi)= 2 e^{-\chi}\int^{\chi}_{-\infty} d\chi'\,e^{\chi' - \int^{\chi}_{\chi'}d\chi''\,\left(1 + \frac{3}{2}\frac{\Omega_m(\chi'')}{f_+^2(\chi'')}\mu(\chi'')\right)}\,.
\label{C22nDGP}
\ee
Analogously to the $\Lambda$CDM case we can find the differential equations and the analytic solutions for the third order kernel and parameters: the result of these calculations are presented in App.~\ref{ansolsn}.
 \begin{figure}[t]
\centering 
\includegraphics[width=.43\textwidth,clip]{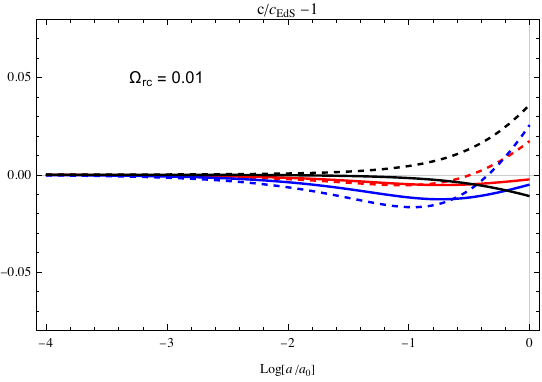}
\includegraphics[width=.53\textwidth,clip]{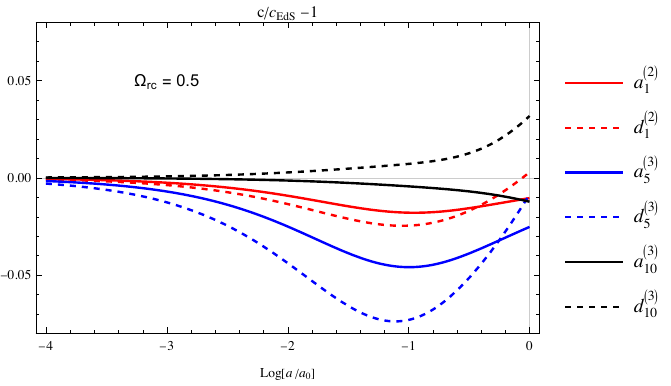}
\caption{Evolution of the coefficients of the kernels in nDGP (for $\Omega_m^0=0.27$ and two possible values of the nDGP parameter $\Omega_{rc}  = 0.01,\,0.5$ in the left and right plot, respectively),  normalized to the EdS limit.}
\label{coeffsn}
\end{figure}
An example of the behaviour of the coefficients in a nDGP cosmology is presented in Fig.~\ref{coeffsn}. Notice that, similarly to what happened in the $\Lambda$CDM case, the velocity coefficients (dashed lines) are more sensible to a change in the time behavior of the growing mode compared to the matter ones (solid lines).
A comparison between the exact solutions in the $\Lambda$CDM case and the nDGP one is given in Fig.~\ref{coeffsr}, for two representative values of the nDGP parameter $\Omega_{rc}  $.
 \begin{figure}[t]
\centering 
\includegraphics[width=.43\textwidth,clip]{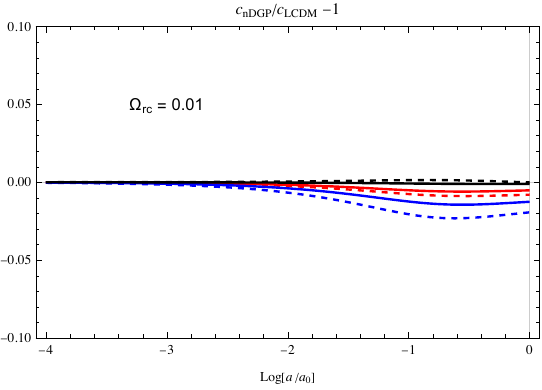}
\includegraphics[width=.53\textwidth,clip]{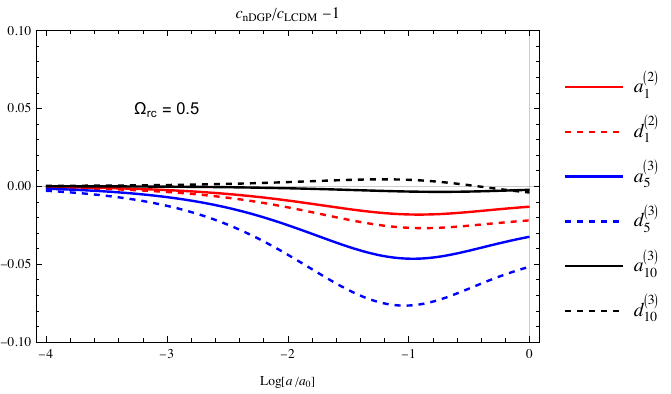}
\caption{Comparison of the time evolution of the matter kernel coefficients in nDGP cosmologies compared to the $\Lambda$CDM ones ($\Omega_m^0=0.27$ and $\Omega_{rc}  = 0.01,\,0.5$ in the left and right plot, respectively).}
\label{coeffsr}
\end{figure}

\section{General tracers}
\label{gentrac}

\subsection{Kernels}


The number density contrast for a generic tracer is not expected to satisfy a continuity equation, or to fulfil momentum conservation. Therefore, in order to obtain the most generic nonlinear kernels for biased tracers, the condition  from mass and momentum conservation given by eq.~\re{km} should be lifted, while keeping the ones  from EGI and the Equivalence Principle. 

The explicit calculations of the kernels in this case can be also found in App.~\ref{app:kernels}. Up to third order, mass and momentum conservation  give four independent constraints. So, compared to the matter case, which is described by three independent coefficients (one for $n=2$ and two for $n=3$), the kernels for general tracers have a total of 7 independent coefficients, which can be chosen to be
\be
\left\{\CC_{0}^{(1)} ,\; \CC_{0}^{(2)},\; \CC_{1}^{(2)},\; \CC_{0}^{(3)},\; \CC_{1}^{(3)}, \;  \CC_{5}^{(3)},\;\CC_{10}^{(3)}\right\}\,.
\label{indpar}
\ee
The kernels are  given by,
\begin{align}
\Kone (\bq_1) & = \CC_{0}^{(1)}\,,\label{Kone}  \\
\Ktwo(\bq_1,\bq_2) & = \,\CC_{0}^{(2)}+2\, \CC_{0}^{(1)} \,\beta(\bq_1,\bq_2)+  \CC_{1}^{(2)} \, \gamma(\bq_1,\bq_2) \,, \label{Ktwo} \\
\Kthree(\bq_1,\bq_2, \bq_3) &= \frac{1}{3}\, \CC_{0}^{(3)} + \CC_{1}^{(3)} \gamma(\bq_1,\bq_2) + 2  \CC_{0}^{(2)}  \beta(\bq_1,\bq_2) \nonumber \\
& +\CC_{5}^{(3)}   \gamma(\bq_1,\bq_2)  \gamma(\bq_{12},\bq_3)   + 2\, \CC_{0}^{(1)} \beta(\bq_1,\bq_2)\beta(\bq_{12},\bq_3) \nonumber \\
&+ 2 ( \hh \, \CC_{0}^{(1)} -\,\CC_{10}^{(3)} ) \gamma(\bq_1,\bq_2)\beta(\bq_{12},\bq_3)  + 2 (\CC_{1}^{(2)}  + 2\,\CC_{10}^{(3)} -  \hh \,\CC_{0}^{(1)})\beta(\bq_1,\bq_2)\gamma(\bq_{12},\bq_3)\nonumber\\
& + \CC_{10}^{(3)} \gamma(\bq_1,\bq_2)\alpha_a(\bq_{12},\bq_3) + {\rm cyclic}\,,
\label{generalK}
\end{align}
where the coefficient $\hh$ has been defined in eq.~\re{hhdef}. 
Notice that it enters the matter kernels  and depends only on the underlying cosmology   and not on the type of tracer. For instance, in the EdS case, it is given by (see eq.~\re{EdS2})
\beq
h_{\rm EdS}(\eta) = \frac{3}{7}\,.
\eeq 
The fact that it appears explicitly  in the tracer kernels opens the possibility, at least in principle, to extract cosmological information in a model independent way.

 \subsection{Relation with other bias expansions}
 \label{sectbias}
  
Here we compare  the present approach to other  bias expansions (see \cite{Desjacques:2010gz,Desjacques:2016bnm} for a review on bias). 
For instance, we can compare our general kernel expansion up to third order in Eqs.~\eqref{Kone}, \eqref{Ktwo} and \eqref{generalK}, with the bias expansion given in  
 \cite{Eggemeier:2018qae}.
Up to third order in PT,  the density contrast for a given biased tracer in configuration space is expressed as the sum of 7 independent operators,
\begin{align}
\delta_{t} =b_1\,\delta +\frac{b_2}{2}\,\delta ^2+\frac{b_3}{3!}\,\delta ^3+b_{{\cal G}_2}\,{\cal G}_2(\Phi)+b_{{\cal G}_3}\,{\cal G}_3(\Phi)+b_{\delta{\cal G}_2}\,\delta \; {\cal G}_2(\Phi)+b_{{\cal G}_N}\,{\cal G}_N(\varphi_2,\varphi_1) \; ,
\label{bexp}
\end{align}
where we have omitted the time dependence and we have defined the two Galilean-invariant combinations
\be
\begin{split}
{\cal G}_2(\Phi)&\equiv\left(\nabla_{ij} \Phi \right)^2-\left(\nabla^2 \Phi \right)^2\,,\\
{\cal G}_3(\Phi)&\equiv\left(\nabla^2 \Phi \right)^3+2\nabla_{ij}\Phi \nabla_{jk}\Phi \nabla_{ki}\Phi -3\left(\nabla_{ij} \Phi \right)^2\nabla^2 \Phi \,,
\end{split}
\ee
where $\Phi$ is the Poisson potential normalized in such a way that $\nabla^2 \Phi = \delta$, and the  Galilean-invariant ``non-local'' combination
\beq
{\cal G}_N(\varphi_2,\varphi_1)  \equiv \nabla_{ij}\varphi_2 \nabla_{ij}\varphi_1  - \nabla^2\varphi_2  \nabla^2\varphi_1 \,,
\eeq
with $\nabla^2\varphi_1 \equiv -\delta$ and $\nabla^2\varphi_2 \equiv - {\cal G}_2(\Phi) $.

The number of independent bias coefficients is the same as ours (cf.~with eq.~\eqref{indpar}): 
\begin{align}
\text{Our basis:}&  \quad  \text{$1^{\rm st}$ order: } \CC^{(1)}_0 \;, \qquad \text{$2^{\rm nd}$ order: } \CC^{(2)}_0, \, \CC^{(2)}_0 \;, \qquad \text{$3^{\rm rd}$ order: } \CC^{(3)}_0, \, \CC^{(3)}_1 , \, \CC^{(3)}_5, \, \CC^{(3)}_{10} \;,  \nonumber \\
\text{Ref.~\cite{Eggemeier:2018qae}:}&  \qquad \text{$1^{\rm st}$ order: } b_1 \;, \qquad \text{$2^{\rm nd}$ order: } b_2, \, b_{{\cal G}_2} \;, \qquad \text{$3^{\rm rd}$ order: } b_3, \, b_{{\cal G}_3} , \, b_{\delta {\cal G}_2}, \, b_{{\cal G}_N} \;.   \nonumber
\end{align}
Indeed, by expanding the RHS of eq.~\eqref{bexp} and 
equating the coefficients of the independent operators, we can relate them as
\be
\begin{split}
\CC^{(1)}_{0}&=b_1,\qquad \CC^{(2)}_{0}=b_2,\qquad \CC^{(3)}_{0} =b_3 , \\ \qquad 
\CC^{(2)}_{1}& =b_1 \, \AC^{(2)}_{1} - 2 \,b_{{\cal G}_2} \,, \qquad \CC^{(3)}_{1}=b_2\, \AC^{(2)}_{1}- 2 \,b_{\delta{\cal G}_2},\\
\CC^{(3)}_{5}& =b_1\, \AC^{(3)}_{5}- 2 \,b_{{\cal G}_2} \AC^{(2)}_{1} + 2\,b_{{\cal G}_3} + 2\,b_{{\cal G}_N},  \qquad \CC^{(3)}_{10}=b_1 \, \AC^{(3)}_{10}- \,b_{{\cal G}_3} \;,
\end{split}
\ee
or, inversely,
\be
\begin{split}
b_1&= \CC^{(1)}_0 ,\qquad  b_2= \CC^{(2)}_0,  \qquad b_3  = \CC^{(3)}_{0}, \\ 
b_{{\cal G}_2} & = \frac12 \left( \AC^{(2)}_1 \CC^{(1)}_0 - \CC^{(2)}_1 \right)  \,, \qquad b_{\delta {\cal G}_2}  = \frac12 \left( \AC^{(2)}_1 \CC^{(2)}_0 - \CC^{(3)}_1 \right)   ,\\
b_{  {\cal G}_3}  & =  \AC^{(3)}_{10} \CC^{(1)}_0 - \CC^{(3)}_{10},  \qquad  b_{  {\cal G}_N}  = \frac12 \left( \left( \AC^{(2)}_1\right)^2 \CC^{(1)}_0 - \AC^{(2)}_1 \CC^{(2)}_1  - \AC^{(3)}_5 \CC^{(1)}_0 + \CC^{(3)}_5 \right) -  \AC^{(3)}_{10} \CC^{(1)}_0 +  \CC^{(3)}_{10}   \;.
\end{split}
\ee
Other   basis expansions  at this order have been given, for instance, in \cite{Chan:2012jj,Saito:2014qha,Assassi:2014fva,Mirbabayi:2014zca,Senatore:2014eva,Angulo:2015eqa,Fujita:2016dne}.
A comparison shows that, for a fixed cosmology, that is, for fixed $a^{(2,3)}_i$ coefficients, our basis can be related in a similar way as above  to those in these references, (see also  \cite{Fujita:2020xtd, Donath:2020abv} for explicit relations between  coefficients). 

For $\Lambda$CDM cosmology, our bias expansion can also be compared to the one presented in \cite{Donath:2020abv}.  Notice that the time-dependent function $Y$ defined in that reference, which carries information about the exact time dependence away from the EdS case, is related to our function $h$ defined in eq.~\re{hhdef}.  More precisely we have: $Y=h/2-3/14$. However, our definition in  eq.~\re{hhdef} is not restricted to $\Lambda$CDM. It applies to any cosmological model sharing the same symmetries as $\Lambda$CDM. Moreover, our derivation clarifies the physical origin of the tracer-independent function $h$, that is, EGI.

\subsection{Relation with Fujita \& Vlah}
Before closing this section, we discuss in some detail the relation of the present approach to that of \cite{Fujita:2020xtd}, which is the closest one to ours. The starting point, also in that paper, is to write down the most general kernels and then to reduce the number of independent coefficients by imposing symmetry-related constraints. In particular, the authors impose that the correlators satisfy the equal-time consistency relations \cite{Peloso:2013zw, Kehagias:2013yd}. On top of that, they also impose extra conditions on the momentum dependence of the soft limit of the four-point function that do not derive directly from the equal time consistency relations.
At third order and for a generic tracer they find 7 bias coefficients and 2 `universal' ones, which are independent on the tracer type and depend on the underlying cosmology only. We have shown that EGI at LO fixes one of  these universal coefficients and at NLO relates the other to the $h(\eta)$ quantity defined in \re{hhdef}. 

Let us see this in detail. Ref.~\cite{Fujita:2020xtd} implements the requirement that the soft limit of the equal time mixed  three point function
\beq
\lim_{q\to 0} \langle \delta_\bq(\eta) \delta_{t_1, \bk_+}(\eta) \delta_{t_2, \bk_-}(\eta) \rangle\,,
\eeq 
where $\bk_\pm\equiv \bk\pm \bq/2$, and `$t_1$' and `$t_2$' indicate two different tracers, has no poles in $q$.
Eqs.~\re{Kone} and \re{kappa2appB} give,
\beq
\lim_{q\to 0} \langle \delta_\bq(\eta) \delta_{t_1, \bk_+}(\eta) \delta_{t_2, \bk_-}(\eta) \rangle = -\frac{1}{2}\frac{\bk\cdot\bq}{q^2}\left( \CC_{2,t_1}^{(2)}(\eta) \CC_{0,t_2}^{(1)}(\eta)  - \CC_{2,t_2}^{(2)}(\eta) \CC_{0,t_1}^{(1)}(\eta) \right) P_l(q;\eta)P_l(k;\eta)\,,
\eeq
where $P_l(q;\eta)$ is the linear power spectrum. Requiring the absence of poles one finds
\beq
\label{cunidef}
\frac{\CC_{2,t_1}^{(2)}(\eta)}{ \CC_{0,t_1}^{(1)}(\eta)}=\frac{\CC_{2,t_2}^{(2)}(\eta)}{ \CC_{0,t_2}^{(1)}(\eta)}\equiv 2 \,{\cal C}_b(\eta)\,,
\eeq
where the quantity  ${\cal C}_b(\eta)$, the first `universal' coefficient defined in  ref.~\cite{Fujita:2020xtd}, does not depend on the tracer type. The authors also observe that in the EdS case ${\cal C}_b(\eta)=1$. From eq.~\re{B21},  this result is completely general, i.e., not limited to EdS. It is a consequence of the EGI of the system. 

This can also be seen by extending the approach of \cite{Fujita:2020xtd}  by using unequal-time consistency relation, which reads \cite{Peloso:2013zw},
\beq
\lim_{q\to 0} \langle \delta_\bq(\eta) \delta_{t_1, \bk_+}(\eta_1) \delta_{t_2, \bk_-}(\eta_2) \rangle =  -\frac{\bk\cdot\bq}{q^2} \left(\frac{D_+(\eta_2)}{D_+(\eta)}-\frac{D_+(\eta_1)}{D_+(\eta)}\right) P_l(q;\eta)P_{t_1,t_2}(k;\eta_1,\eta_2)\,,
\label{unCR}
\eeq
where $P_{t_1,t_2}(k;\eta_1,\eta_2)$ is the unequal time cross PS between the tracers $t_1$ and $t_2$. Using again eqs.~\re{Kone} and \re{kappa2appB} we get
\begin{align}
\lim_{q\to 0}& \langle \delta_\bq(\eta) \delta_{t_1, \bk_+}(\eta_1) \delta_{t_2, \bk_-}(\eta_2) \rangle = \nonumber\\
& \qquad\qquad-\frac{1}{2}\frac{\bk\cdot\bq}{q^2} \left(\frac{D_+(\eta_2)}{D_+(\eta)} \frac{\CC_{2,t_2}^{(2)}(\eta_2)}{ \CC_{0,t_2}^{(1)}(\eta_2)}-
\frac{D_+(\eta_1)}{D_+(\eta)} \frac{\CC_{2,t_1}^{(2)}(\eta_1)}{ \CC_{0,t_1}^{(1)}(\eta_1)} \right) P_l(q;\eta)P_{t_1,t_2}(k;\eta_1,\eta_2)\,,
\end{align}
which reproduces the consistency relation \re{unCR} if 
\beq
\frac{\CC_{2,t_1}^{(2)}(\eta)}{ \CC_{0,t_1}^{(1)}(\eta)}=\frac{\CC_{2,t_2}^{(2)}(\eta)}{ \CC_{0,t_2}^{(1)}(\eta)}= 2 \,,
\eeq
thus confirming ${\cal C}_b=1$ (see eq.~\re{cunidef}) in any cosmology respecting the equivalence principle.

On the other hand, the other universal coefficient, ${\cal C}_d$,  is indeed cosmology-dependent. By using the NLO condition \re{nlo}, on the third order kernel  in eq.~(C.1) of \cite{Fujita:2020xtd}, one can verify that 
\beq
{\cal C}_d(\eta) = 2\, h(\eta)\,,
\eeq
where the non-local in time, tracer-independent coefficient $h(\eta)$ is given in \re{hhdef}.

 \section{UV effects}
 \label{sec:HD}
Up to this point, we have implicitly assumed that PT is able to model the nonlinear behavior at all scales. As it is well known, this assumption fails at small scales,  for wavenumbers larger than some value $k_\mathrm{NL}$.
More precisely, the effect of ``UV'' modes with $q>k_\mathrm{NL}$ on the nonlinear field of a given tracer evaluated at $k<k _\mathrm{NL}$ is poorly reproduced by PT, see for instance \cite{Nishimichi:2014rra}.
In fact, the physics at small scales is unknown and its effects need to be taken into account as an expansion in $k/k_{\rm NL}$ \cite{Carrasco:2012cv}.

In order to keep track of the UV corrections needed to correct PT up to a given order, we perform the shift
\be
\vphi_{\bq}(\eta) \to \vphi_{\bq}(\eta) + \delta \vphi_{\bq}(\eta)\,,
\label{51}
\ee
in the expansions \re{expansion_d2},  \re{expansion_t2}, or  \re{expansion_v2},
where
\be
 \delta \vphi_{\bq}(\eta) \equiv \left(\tilde\vphi_\bq(\eta) -\vphi_\bq(\eta) \right) \Theta(q-\Lambda)\equiv  \tilde\vphi^{\rm{uv}}_{\bq}(\eta) -\vphi^{\rm{uv}}_{\bq}(\eta) \,.
 \label{52}
\ee
$\tilde\vphi^{\rm{uv}}_{\bq}(\eta)$ is the `true' UV field which replaces the ``wrong'' one,  $\vphi^{\rm{uv}}_{\bq}(\eta)$ so that, once inserted in the loop corrections, the wrong PT behavior in the UV is replaced by  the corrected one.
 In the following, we will assume the hierarchy $\Lambda \gg k_{\rm{NL}} \gg k$, where $k$ is the momentum associated to the nonlinear field. In this limit, the UV effects can be expressed as  powers of $k^2/k^2_{\rm{NL}}$, and the $\Lambda$ dependence can be omitted\footnote{The regularization scale $\Lambda$ is non-physical. Therefore the fully renormalized PT observables should not depend on it.}, as we did in \re{51} and in the last of \re{52}.

Under the shift \re{51} the nonlinear expansion of the tracer field, eq.~\re{expansion_v2}, gets extra contributions,
\begin{align}
\delta^{(n)}_{t,\bk}(\eta) &= \Ic_{\bk;\bq_1\cdots,\bq_n}\, \Kn(\bq_1,\cdots,\bq_n;\eta) \vphi_{\bq_1}(\eta)\cdots  \vphi_{\bq_n}(\eta),\nonumber\\
&\rightarrow  \sum_{m=0}^n \left(\begin{array}{c} n\\m\end{array}\right) \Ic_{\bk;\bq_1\cdots,\bq_n}\, \Kn(\bq_1,\cdots,\bq_n;\eta) \delta \vphi_{\bq_1}(\eta)\cdots  \delta \vphi_{\bq_m}(\eta)\vphi_{\bq_{m+1}}(\eta)\cdots \vphi_{\bq_n}(\eta)\,,
\label{expansion_UV}
\end{align}
where the $m > 0$ terms correct the wrong UV behavior of the $m=0$ one. We now discuss the effect of these corrections up to $n=3$.
For $n=1$ we have no modification,
\be
\delta^{(1)}_{t,\bk}(\eta) \to \Kone(\bk;\eta)\left( \vphi_{\bk}(\eta)+\delta \vphi_{\bk}(\eta)\right) = \Kone(\bk;\eta)\vphi_{\bk}(\eta)  \,,
\ee
as, within the assumed hierarchy of scales, the UV field $\delta \vphi_{\bk}(\eta)$ has no support on the IR momentum $k$.
For $n=2$, 
\begin{align}
\delta^{(2)}_{t,\bk}(\eta) \to & \Ic_{\bk;\bq_1,\bq_2}\, \Ktwo(\bq_1,\bq_2;\eta)( \vphi_{\bq_1}(\eta)+\delta  \vphi_{\bq_1}(\eta))( \vphi_{\bq_2}(\eta)+\delta  \vphi_{\bq_2}(\eta))
\nonumber\\
=& \, \Ic_{\bk;\bq_1,\bq_2}\, \Ktwo(\bq_1,\bq_2;\eta) \vphi_{\bq_1}(\eta)  \vphi_{\bq_2}(\eta)\nonumber\\
&+ \Ic_{\bk;\bq_1,\bq_2}\, \Ktwo(\bq_1,\bq_2;\eta) \left(\tilde\vphi^{\rm{uv}}_{\bq_1}(\eta) \tilde \vphi^{\rm{uv}}_{\bq_2}(\eta) -\vphi^{\rm{uv}}_{\bq_1}(\eta)  \vphi^{\rm{uv}}_{\bq_2}(\eta)  \right)\,,
\label{55}
 \end{align}
 where in the last equation we have again made use of the hierarchy of scales. 
the new term contains only UV fields, and, for the matter and velocity fields, in the small $k$ limit, it  scales as $k^2/k_{\rm{NL}}^2$, due to momentum conservation. The full correction to the second order field will then have the form, 
 \be
 \Ic_{\bk;\bq_1,\bq_2}\, \Ktwo(\bq_1,\bq_2;\eta) \left(\tilde\vphi^{\rm{uv}}_{\bq_1}(\eta) \tilde \vphi^{\rm{uv}}_{\bq_2}(\eta) -\vphi^{\rm{uv}}_{\bq_1}(\eta)  \vphi^{\rm{uv}}_{\bq_2}(\eta)  \right)\stackrel{k \to 0}{\longrightarrow} \frac{k^2}{k_{\rm{NL}}^2} \epsilon^{(2)}_\delta(\eta)\,,
 \ee
 where the zero average stochastic field $\epsilon^{(2)}_\delta(\eta)$ will be assumed to be uncorrelated with $\vphi_{\bq}(\eta)$.

On the other hand, for a general tracer there is no momentum constraint, and the new term behaves as a scale independent white noise for small $k$. Therefore, we may write
 \begin{align}
 \delta_{\bk}^{(2)} (\eta)& =  \delta_{\bk}^{(2),\,PT}(\eta) + \frac{k^2}{k_{\rm{NL}}^2} \epsilon^{(2)}_\delta(\eta) +O\left( \frac{k^4}{k_{\rm{NL}}^4}\right)\,,\nonumber\\
 \theta_{\bk}^{(2)} (\eta)& =  \theta_{\bk}^{(2),\,PT}(\eta) + \frac{k^2}{k_{\rm{NL}}^2} \epsilon^{(2)}_\theta(\eta) +O\left( \frac{k^4}{k_{\rm{NL}}^4}\right)\,,\nonumber\\
\delta_{t,\bk}^{(2)} (\eta)& =  \delta_{t,\bk}^{(2),\,PT}(\eta) +\epsilon^{(2)}_t(\eta) +O\left( \frac{k^2}{k_{\rm{NL}}^2}\right)\,,
\label{noise2}
 \end{align}
 where by $\delta_{\bk}^{(2),\,PT}(\eta)$ and so on we indicate the uncorrected results of the previous sections, and we define the stochastic fields $ \epsilon^{(2)}_{\delta,\theta,t}$ from the $k\to 0$ limit of the second line in eq.~\re{55}.
 
At $n=3$ we get
\begin{align}
\delta^{(3)}_{t,\bk}(\eta) \to &  \delta_{t,\bk}^{(3),\,PT}(\eta) \nonumber\\
& +  \Ic_{\bk;\bq_1,\bq_2,\bq_3}\, \Kthree(\bq_1,\bq_2,\bq_3;\eta) \bigg[ 3 \;\vphi^{\rm{ir}}_{\bq_1}(\eta) \left(\tilde\vphi^{\rm{uv}}_{\bq_2}(\eta) \tilde \vphi^{\rm{uv}}_{\bq_3}(\eta) -\vphi^{\rm{uv}}_{\bq_2}(\eta)  \vphi^{\rm{uv}}_{\bq_3}(\eta)  \right)\nonumber\\
&\qquad\qquad\qquad \quad+\left(  \tilde\vphi^{\rm{uv}}_{\bq_1}(\eta) \tilde\vphi^{\rm{uv}}_{\bq_2}(\eta) \tilde \vphi^{\rm{uv}}_{\bq_3}(\eta) - \vphi^{\rm{uv}}_{\bq_1}(\eta) \vphi^{\rm{uv}}_{\bq_2}(\eta)  \vphi^{\rm{uv}}_{\bq_3}(\eta)    \right)\bigg]\,,
\label{57}
\end{align}
where $\vphi^{\rm{ir}}_{\bq}(\eta)\equiv \vphi_{\bq}(\eta)\Theta(\Lambda-q)$. The last line gives again noise terms uncorrelated with $ \vphi_{\bq}(\eta)$ and with the momentum dependences of eq.~\re{noise2}. On the other hand, the second line contains the IR component of the linear field. By taking the functional derivative of  $\delta^{(3)}_{t,\bk}(\eta)$ with respect to $\vphi^{\rm{ir}}_{\bk}(\eta)$, this term gives 
\begin{align}
\frac{\delta\; \delta^{(3)}_{t,\bk}(\eta)}{ \delta\; \vphi^{\rm{ir}}_{\bk}(\eta)} =\cdots+ \frac{3}{(2 \pi)^3} \int\frac{d^3 q}{(2\pi)^3} \Kthree(\bk,\bq,-\bq;\eta) \left(\tilde\vphi^{\rm{uv}}_{\bq}(\eta) \tilde \vphi^{\rm{uv}}_{-\bq}(\eta) -\vphi^{\rm{uv}}_{\bq}(\eta)  \vphi^{\rm{uv}}_{-\bq}(\eta)  \right)\,,
\label{58}
\end{align}
which, again, vanishes as $k^2/k_{\rm{NL}}^2$ for matter density and velocity, and goes to a constant for a generic tracer as $k\to 0$. The expectation value of the RHS of \re{58}  for matter and velocity corresponds to the `sound speed' of the EfToLSS \cite{Carrasco:2012cv,Manzotti:2014loa,Baldauf:2015aha,Noda:2017tfh},
\begin{align}
c_{s,\delta}^2(\eta) = \lim_{k\to 0} \frac{k^2_{\rm{NL}}}{k^2} \left\langle \frac{\delta\; \delta^{(3)}_{\bk}(\eta)}{ \delta\; \vphi^{\rm{ir}}_{\bk}(\eta)}  \right\rangle\,,\qquad c_{s,\theta}^2(\eta) = \lim_{k\to 0} \frac{k^2_{\rm{NL}}}{k^2} \left\langle \frac{\delta\; \theta^{(3)}_{\bk}(\eta)}{ \delta\; \vphi^{\rm{ir}}_{\bk}(\eta)}  \right\rangle\,,
\end{align}
while, for a generic tracer it gives both a contribution degenerate with linear bias,  at $O(k^0)$, and a sound speed one at $O(k^2)$. In summary, the $n=3$ result of eq.~\re{57} can be expressed as 
 \begin{align}
 \delta_{\bk}^{(3)} (\eta)& =  \delta_{\bk}^{(3),\,PT}(\eta) + \frac{k^2}{k_{\rm{NL}}^2} \left[ \left( c_{s,\delta}^2(\eta)+ \eta^{(3)}_{\delta}(\eta)\right) \vphi_{\bk}(\eta)+ \epsilon^{(3)}_\delta(\eta)\right] +O\left( \frac{k^4}{k_{\rm{NL}}^4}\right)\,,\nonumber\\
 \theta_{\bk}^{(3)} (\eta)& =  \theta_{\bk}^{(3),\,PT}(\eta) + \frac{k^2}{k_{\rm{NL}}^2} \left[ \left( c_{s,\theta}^2(\eta)+ \eta^{(3)}_{\theta}(\eta)\right) \vphi_{\bk}(\eta)+ \epsilon^{(3)}_\theta(\eta)\right] +O\left( \frac{k^4}{k_{\rm{NL}}^4}\right)\,,\nonumber\\
\delta_{t,\bk}^{(3)} (\eta)& =  \delta_{t,\bk}^{(3),\,PT}(\eta) + \left[ \tilde b_{0,t}(\eta)+ \eta^{(3)}_{t}(\eta)+ c_{s,t}^2(\eta)  \frac{k^2}{k_{\rm{NL}}^2} \right] \vphi_{\bk}(\eta)+ \epsilon^{(3)}_\theta(\eta)+O\left( \frac{k^2}{k_{\rm{NL}}^2}\right)\,,
\label{noise3}
 \end{align}
 where the stochastic fields $ \eta^{(3)}_{\delta,\theta,t}(\eta)$ are given by the fluctuation of eq.~\re{57} about its expectation value (multiplied by $k^2_{\rm{NL}}/k^2$ in the case of $\delta$ and $\theta$).
 
\section{Conclusions}
\label{concl}
In this paper we have investigated the role of symmetries in determining the analytic structure of PT kernels, both for DM and for generic biased tracers. We have highlighted the prominent role played by EGI, and showed that  the constraints imposed by this symmetry exhibit a rich structure. The role of EGI was already well appreciated in the literature \cite{Fujita:2020xtd}, as it is at the basis of the consistency relations for the LSS \cite{Peloso:2013zw,Kehagias:2013yd} (for recent applications see \cite{Marinucci:2019wdb,Marinucci:2020weg}), which relate $N$-point correlators to $N-m$ ones  as $m$ external momenta vanish independently. This limit, and therefore the consistency relations, correspond to the EGI relations  at LO in our language. On the other hand, we showed that the limits in which  partial sums of the external momenta vanish are also governed by EGI, and provide extra  constraints on the functional form of the kernels, as is summarized in eq.~\re{genGIshort}. 

The number of independent EGI constraints  increases with the PT order. At second order, the LO EGI gives one constraint,  while, at third order,  LO+NLO EGI provide eight independent conditions. Going to fourth order, one should include also NNLO constraints, and so on. We leave a systematic discussion of  PT kernels beyond third order to future work. 

Another noticeable feature of eq.~\re{genGIshort} is its non-locality in time, which is made explicit at  third order by the coefficient  $h(\eta)$, defined in eq.~\re{hhdef}. Time non-locality is a well known feature of the bias (and EFT) expansion (for a discussion, see for instance  \cite{Mirbabayi:2014zca,Senatore:2014eva}), where it is related to the convective time derivatives appearing in the equations of motion. It is not surprising that, in our discussion, it emerges as a consequence of EGI, which also provides a systematic way to take it into account at higher orders.

The symmetry-based framework defined in this paper allows for a general treatment of DM and biased tracers, as the two species differ only in mass and momentum conservation, which is enforced for the former but not for the latter. It also provides a useful language for using galaxy clustering data for model independent analyses of beyond $\Lambda$CDM scenarios, which could be classified in terms of symmetries rather than specific Lagrangians. 
Considering for instance the galaxy PS in redshift space at one loop, the cosmology-dependent coefficients are $h(\eta)$ and the ones appearing in the velocity kernels, $d_1^{(2)}$, $d_5^{(3)}$, and $d_{10}^{(3)}$. They can either be computed for a given model, as we did for $\Lambda$CDM and for nDGP in sect.~\ref{time-dep}, or fit from data in a model independent analysis.

It was shown in \cite{Crisostomi:2019vhj} that in Horndeski theories EGI symmetry holds and the structure of the PT kernels is the same as that in $\Lambda$CDM. 
Beyond Horndeski (see e.g.~\cite{Zumalacarregui:2013pma,Gleyzes:2014dya,Langlois:2015cwa}), EGI symmetry is broken and PT kernels and the bias expansion deviate from the standard ones, leading for instance to violations of the consistency relations (see \cite{Lewandowski:2019txi} for a thorough study of the consequences of this on observables). It would be interesting to systematically study these deviations in a general setting and constrain them with data.
A challenging direction is the extension of this approach to modified gravity models involving a scale-dependent growth, such as for instance $f(R)$ theories, which we leave for the future.

\vskip 0.5 cm

\centerline{ \bf\Large{Acknowledgments}}
\vskip 0.2 cm
MP acknowledges support from the research grant ``The Dark Universe: A Synergic Multimessenger Approach No. 2017X7X85'', funded by MIUR.

\appendix

\section{Galilean invariance for an arbitrary time dependence}
\label{A1}

In this Appendix we show that the constraints from EGI do not require to assume a particular time dependence for the displacement defining the velocity shift, as we did in eq.~\re{tdeq}.
Since we are going to have arbitrary time dependence, we have to allow different times for the linear fields in the expansion for $\delta_{t,\bk}(\eta)$ in eq.~\re{expansion}, namely,
 \begin{align}
  \delta_{t,\bk}(\eta)& =_{p.t.} \, \sum_{n=1}^\infty \frac{1}{n!}\int \frac{d^3 q_1}{(2\pi)^3} \cdots \frac{d^3 q_n}{(2\pi)^3}\nonumber\\
  &\qquad\times\int^\eta_{\eta_{in}}d\eta_1\cdots \int^\eta_{\eta_{in}}d\eta_n \left.\frac{\delta^n  {\cal F}_{t}[\vphi^{\lambda}_\bq(\eta)](\bk,\eta)}{\delta \vphi^{\lambda_1}_{\bq_1}(\eta_1)\cdots \delta \vphi^{\lambda_n}_{\bq_n}(\eta_n)}\right|_{\vphi^{\lambda}_\bq(\eta)=0} \vphi^{\lambda_1}_{\bq_1}(\eta_1)\cdots  \vphi^{\lambda_n}_{\bq_n}(\eta_n)\nonumber\\
&= \sum_{n=1}^\infty \,\Ic_{\bk;\bq_1\cdots,\bq_n}\,\int^\eta_{\eta_{in}}d\eta_1 \cdots \int^\eta_{\eta_{in}}d\eta_n \nonumber\\
&\qquad\times \hat \Kn_{\lambda_1\cdots\lambda_n}(\bq_1,\cdots,\bq_n;\eta;\eta_1,\cdots,\eta_n)f^+_{\bq_1}(\eta_1)\cdots f^+_{\bq_n}(\eta_n) \vphi^{\lambda_1}_{\bq_1}(\eta_1)\cdots  \vphi^{\lambda_n}_{\bq_n}(\eta_n)\,,\nonumber\\
&=\sum_{n=1}^\infty \,\Ic_{\bk;\bq_1\cdots,\bq_n}\, \Kn(\bq_1,\cdots,\bq_n;\eta) \vphi_{\bq_1}(\eta)\cdots  \vphi_{\bq_n}(\eta)\,,
\label{expansion2}
 \end{align}
 where now
 \begin{align}
 &\Kn(\bq_1,\cdots,\bq_n;\eta) \nonumber\\
& = \int^\eta_{\eta_{in}}d\eta_1 \cdots \int^\eta_{\eta_{in}}d\eta_n  \hat \Kn_{\lambda_1\cdots\lambda_n}(\bq_1,\cdots,\bq_n;\eta;\eta_1,\cdots,\eta_n)f^+_{\bq_1}(\eta_1)\cdots f^+_{\bq_n}(\eta_n)\nonumber\\
&\qquad\qquad \qquad\qquad\qquad\qquad \times\frac{D_{\bq_1}(\eta_1)}{D_{\bq_1}(\eta)}\cdots \frac{D_{\bq_n}(\eta_n)}{D_{\bq_n}(\eta)}\,u^{\lambda_1}_{f,\bq_1}(\eta_1)\cdots u^{\lambda_n}_{f,\bq_n}(\eta_n)\,.
\label{KKh}
 \end{align}
 Now, let us consider again a shift of the form
 \beq
 \vphi^{\lambda_i}_{\bq_i}(\eta_i)\to  \vphi^{\lambda_i}_{\bq_i}(\eta_i) +i \delta_{\lambda 2}(2\pi)^3\delta_D(\bq_i) \,  \frac{\bq_i\cdot {\dot\bd}(\eta_i)}{f_{\bq_i}^+(\eta_i) {\cal H}(\eta_i)}\,,
 \eeq
 but with an arbitrary time dependence for $\bd(\eta)$. 
 Now, if the soft limit of the multi-time kernels satisfies
 \begin{align}
 \lim_{\bq_1 \cdots\bq_m\to 0}& q_1^{i_1}\cdots q_m^{i_m} \hat \Kn_{\lambda_1\cdots\lambda_n}(\bq_1,\cdots,\bq_n;\eta;\eta_1,\cdots,\eta_n)\nonumber\\
 &=Q_{n,m}^{i_1}\cdots Q_{n,m}^{i_m} \hat K_{n-m, \lambda_{m+1}\cdots\lambda_n}(\bq_{m+1},\cdots,\bq_n;\eta;\eta_{m+1},\cdots,\eta_n)\,,
 \label{KK2}
 \end{align}
 When inserted in eq.~\re{expansion2}, each shift comes with a time integral of the form
  \beq
  \int^\eta_{\eta_{in}}d\eta_i  \,f_{\bq_i=0}^+(\eta_i) \,  \frac{ {\dot d}(\eta_i)}{f_{\bq_i=0}^+(\eta_i) {\cal H}(\eta_i)} = d^i(\eta), \eeq
 (where we have set $d^i(\eta_{in})=0$), which reproduces the $\bk\cdot \bd(\eta)$ factors in \re{exppert}. Using \re{KK2} in \re{KKh}, the EGI constraint of eq.~\re{GIc} is recovered. 
 
 \section{Next-to-Next-to-Leading order}
\label{nnlosec}
In what follows we present a calculation of the Next-to-Next-to-Leading order (NNLO) contribution from the EGI. Going to this order implies that we consider a displacement field produced by the coupling of three linear modes, $\vphi_{\bq_1}$, $\vphi_{\bq_2}$ and $\vphi_{\bq_3}$, given by
\beq
\tilde{\bd}_\bq^{(3)}(\eta) = -i\frac{\bq}{q^2}\int^{\eta}d\eta'\,f_+(\eta')\left(\frac{D_+(\eta')}{D_+(\eta)}\right)^3{\cal I}_{\bq,\bq_1,\bq_2,\bq_3}\Gthree (\bq_1,\bq_2,\bq_3;\eta')\vphi_{\bq_1}(\eta)\vphi_{\bq_2}(\eta)\vphi_{\bq_3}(\eta),
\label{thirddisp}
\eeq
and the zero momentum displacement
\beq
\bd^{(3)}(\eta) = \frac{1}{V}\int_Vd^3\bx\int\frac{d^3\bq}{(2\pi)^3}e^{-i\bq\cdot\bx}\tilde{\bd}_\bq^{(3)}(\eta).
\label{thirdd}
\eeq
The NNLO gives its first contribution to the 4-th order kernel. By expanding eq.~\re{tGI} at fourth order we have
\begin{align}
\label{ftGI}
\delta_{t,\bk}^{(4)}(\eta)\to&\delta_{t,\bk}^{(4)}(\eta)\\
& + i\bk\cdot\bd^{(1)}(\eta)\delta_{t,\bk}^{(3)}(\eta) + i\left(\bk\cdot\bd^{(1)}(\eta)\right)^2\delta_{t,\bk}^{(2)}(\eta) + i\left(\bk\cdot\bd^{(1)}(\eta)\right)^3\delta_{t,\bk}^{(1)}(\eta)\nonumber\\
& + i\bk\cdot\bd^{(2)}(\eta)\delta_{t,\bk}^{(2)}(\eta) + i\bk\cdot\bd^{(1)}(\eta)\,\bk\cdot\bd^{(2)}(\eta)\delta_{t,\bk}^{(1)}(\eta) \nonumber\\
& + i\bk\cdot\bd^{(3)}(\eta)\delta_{t,\bk}^{(1)}(\eta),\nonumber
\end{align}
where we can explicitly see that in the first line of the rhs eq.~\re{ftGI} we have the LO terms which give rise to the relation in eq.~\re{GIc}, while the NLO terms are collected in the second line, and give relations similar to eq.~\re{nlo}.\\
In the third line of eq.~\re{ftGI} it is shown the first NNLO contribution, which will give insights on the behavior of the fourth order kernel when the sum of three momenta goes to zero. This contribution is generated by the couplings of three momenta and is given by 
\begin{align}
\label{GI4}
i\bk\cdot\bd^{(3)}&(\eta)\delta_{t,\bk}^{(1)}(\eta) = \\
&\frac{1}{V}{\cal I}_{0,\bq_1,\bq_2,\bq_3}\frac{\bk\cdot\bq_{123}}{q_{123}^2}\int^{\eta}d\eta'\,f_+(\eta')\left(\frac{D_+(\eta')}{D_+(\eta)}\right)^3 \Gthree(\bq_1,\bq_2,\bq_3;\eta')\Kone(\bk;\eta)\vphi_{\bq_1}(\eta)\vphi_{\bq_2}(\eta)\vphi_{\bq_3}(\eta)\vphi_{\bk}(\eta),\nonumber
\end{align}
and this contribution should be contained in the expression for $\delta_{t,\bk}^{(4)}$. To isolate it we adopt a analogous procedure to that presented in Sec \re{NLOsec}, using the identity
\beq
\label{id4}
1 = \frac{(2\pi)^3}{V}\left[\delta_D(\bq_{123}) + \rm{cyclic} \right] + 1 - \frac{(2\pi)^3}{V}\left[\delta_D(\bq_{123}) + \rm{cyclic}\right],
\eeq
so that
\begin{align}
\delta_{t,\bk}^{(4)}&\supset{\cal I}_{\bk, \bq_1,\bq_2,\bq_3,\bq_4}\Kfour(\bq_1,\bq_2,\bq_3,\bq_4;\eta)\vphi_{\bq_1}(\eta)\vphi_{\bq_2}(\eta)\vphi_{\bq_3}(\eta)\vphi_{\bq_4}(\eta) \frac{(2\pi)^3}{V} \delta_D(\bq_{123})\\
&= \frac{1}{V}{\cal I}_{0,\bq_1,\bq_2,\bq_3}\Kfour(\bq_1,\bq_2,\bq_3,\bk)\vphi_{\bq_1}(\eta)\vphi_{\bq_2}(\eta)\vphi_{\bq_3}(\eta)\vphi_{\bk}(\eta)\nonumber,
\end{align}
which should be set equal to the contribution in eq.~\re{GI4}. We obtain 
\begin{align}
\label{last}
&\lim_{\bq_{123}\to0}\Kfour(\bq_1,\bq_2,\bq_3,\bq_4;\eta)  \supset\\
& \frac{\bk\cdot\bq_{123}}{q_{123}^2}\int^{\eta}d\eta' f_+(\eta')\left(\frac{D_+(\eta')}{D_+(\eta)}\right)^3\Gthree(\bq_1,\bq_2,\bq_3;\eta')\Kone(\bk;\eta) + O\left(\left(\frac{q_{123}}{k}\right)^2\right).\nonumber
\end{align}

\section{Detailed calculation of the  kernels}
\label{app:kernels}

 \subsection{Case $n=2$}
   
 We start from the most generic form of the kernel, 
  \begin{align}
  \Ktwo(\bq_1,\bq_2;\eta) = \,\CC_{0}^{(2)}(\eta)+\CC_{1}^{(2)}(\eta) \,\gamma(\bq_1,\bq_2) +\CC_{2}^{(2)}(\eta) \,\beta(\bq_1,\bq_2)  \,.
  \label{kappa2appB}
 \end{align}
Unless it is necessary, to avoid clutter we suppress the explicit  dependence on $\eta$.
 Now we impose the   constraint eq.~\re{GIc}, which implies
\be
 \begin{split}
 \lim_{\bq_2\to 0}     \Ktwo(\bq_1,\bq_2) = \frac{\bq_1\cdot\bq_2}{q_2^2}  \Kone(\bq_1) =  \frac{\bq_1\cdot\bq_2}{q_2^2} \CC_{ 0}^{(1)} +{\cal O}(q_2^0)
 \,.
 \end{split} 
\ee
Taking the same limit in eq.~\eqref{kappa2appB} we obtain
\be
 \begin{split}
 \lim_{\bq_2\to 0}     \Ktwo(\bq_1,\bq_2) &=\frac{1}{2}\frac{\bq_1\cdot\bq_2}{q_2^2} \CC_{2}^{(2)}    \,,
 \end{split} 
\ee
and comparing the two equations give, as anticipated, a relation between the coefficients of the kernels at different order,
\be
\CC_{2}^{(2)}  =2\, \CC_{ 0}^{(1)} \,.
 \label{B21}
\ee
Therefore, the most generic  kernel at $n=2$ is given by eq.~\eqref{Ktwo}.

For conserved tracers, mass and momentum conservation, eq.~\re{km}, give
\be
 \begin{split}
 &\Ktwo(\bq_1,-\bq_1) = \CC_{ 0}^{(2)} =0\,,\\
 &  \partial_{ q_2^i} \Ktwo(\bq_1,\bq_2) |_{\bq_2=-\bq_1}=0\,,
 \label{momm}
 \end{split}
 \ee
 where, due to the properties of the $\alpha_s$ and $\beta$ functions, the second condition gives no constraint on the coefficients. The most generic matter  kernel at $n=2$ is thus given by eq.~\eqref{kernel2}.

 \subsection{Case $n=3$}
 The most general momentum structure for   $\Kthree(\bq_1,\bq_2,\bq_3)$ is given by\footnote{A contribution of the form $(\bq_1\cdot\bq_2)(\bq_2\cdot\bq_3)(\bq_3\cdot\bq_1)/(q_1^2 q_2^2 q_3^2)$ can be shown to be linearly dependent from those generated by the direct product \re{dirprod}, and therefore needs not to be added.} 
 \begin{align}
\Kthree(\bq_1,\bq_2,\bq_3) =&\,\,\frac{1}{3} \CC_{ 0}^{(3)} + \CC_{ 1}^{(3)} \gamma(\bq_1,\bq_2) +  \CC_{ 2}^{(3)}  \gamma(\bq_{12},\bq_3) + \CC_{ 3}^{(3)} \beta(\bq_1,\bq_2) +  \CC_{ 4}^{(3)} \beta(\bq_{12},\bq_3) \nonumber\\
&+ \CC_{ 5}^{(3)}\gamma(\bq_1,\bq_2)  \gamma(\bq_{12},\bq_3)+  \CC_{ 6}^{(3)} \beta(\bq_1,\bq_2)  \beta(\bq_{12},\bq_3) \nonumber\\
&+ \CC_{ 7}^{(3)} \gamma(\bq_1,\bq_2)  \beta(\bq_{12},\bq_3)+  \CC_{ 8}^{(3)} \beta(\bq_1,\bq_2)  \gamma(\bq_{12},\bq_3)\nonumber\\
&+ \Big( \CC_{ 9}^{(3)}+\CC_{ 10}^{(3)} \gamma(\bq_1,\bq_2) +\CC_{ 11}^{(3)} \beta(\bq_1,\bq_2)  \Big)\,\alpha_a(\bq_{12},\bq_3) \nonumber\\
&+ {\mathrm{ 2 \;permutations}}\,.
\label{k3gappB}
\end{align}
First we apply  eq.~\re{GIc}, in the single vanishing momentum case,
\be
\begin{split}
\lim_{q_1\to 0} \Kthree(\bq_1,\bq_2,\bq_3 ) = \frac{(\bq_2+\bq_3)\cdot\bq_1}{q_1^2}  \Ktwo(\bq_2,\bq_3) \,,
\end{split}
\ee
while in the same limit eq.~\eqref{k3gappB}  gives
\be
\begin{split}
\lim_{q_1\to 0} \Kthree(\bq_1,\bq_2,\bq_3 ) =& \ \frac{(\bq_2+\bq_3)\cdot\bq_1}{q_1^2} \bigg[\frac{1}{2}\left(  \CC_{3}^{(3)} +  \CC_{4}^{(3)} \right)- \CC_{9}^{(3)} +\gamma(\bq_2,\bq_3) \left(   \frac{1}{2} \left( \CC_{7}^{(3)} + \CC_{8}^{(3)} \right)- \CC_{10}^{(3)} \right)\\
&+\beta(\bq_2,\bq_3) \left(   \CC_{6}^{(3)} - \CC_{11}^{(3)} \right)\bigg] + \frac{(\bq_2-\bq_3)\cdot\bq_1}{q_1^2} \CC_{11}^{(3)} \,\alpha_a(\bq_2,\bq_3)  \;.
\end{split}
\ee
Comparing these two equations using the explicit form of $\Ktwo$, eq.~\eqref{Ktwo},
gives the following constraints,
\be
\begin{split}
&\CC_{3}^{(3)}+ \CC_{4}^{(3)}-2 \,\CC_{9}^{(3)}=2 \,\CC_{0}^{(2)}, \qquad   \CC_{7}^{(3)}+\CC_{8}^{(3)} - 2 \CC_{10}^{(3)} =  2 \CC_{1}^{(2)}\,, \qquad \CC_{6}^{(3)}  - \CC_{11}^{(3)}=  \CC_{2}^{(2)}\,, \qquad   \CC_{11}^{(3)}=0\,.
\label{C31}
\end{split}
\ee

The double limit gives
\be
\begin{split}
\lim_{q_1,q_2\to 0} \Kthree(\bq_1,\bq_2,\bq_3) =&\frac{1}{2} \frac{\bq_3\cdot\bq_1}{q_1^2}  \frac{\bq_3\cdot\bq_2}{q_2^2}  \big(   \CC_{6}^{(3)}   - \CC_{11}^{(3)} \big)  = \frac{\bq_3\cdot\bq_1}{q_1^2}  \frac{\bq_3\cdot\bq_2}{q_2^2}  \Kone (\bq_3 ) \,,
\end{split}
\ee
from which we get the (redundant) constraint
\be
\CC_{6}^{(3)} - \CC_{11}^{(3)} = 2\, \CC_{0}^{(1)}\,.
\label{C32}
\ee
Combined with the second and third relations in \re{C31}, this is equivalent to eq.~\re{B21}.

Imposing eq.~\re{nlo} we have
\be
\begin{split}
\lim_{q_{12}\to 0} \Kthree (\bq_1,\bq_2,\bq_3)\,\supset \, \frac{\bq_{12}\cdot\bq_3}{q_{12}^2} \CC_{0}^{(1)} (\eta) \int^\eta d\eta'\, f_+(\eta')\left[\frac{D_+(\eta')}{D_+(\eta)}\right]^2\Gtwo(\bq_1,\bq_2;\eta')
\,,
\end{split}
\ee
while eq.~\eqref{k3gappB} in the same limit gives
\be
\begin{split}
\lim_{q_{12}\to 0} \Kthree (\bq_1,\bq_2,\bq_3) \,\supset & \frac{\bq_{12}\cdot\bq_3}{q_{12}^2} \bigg[\frac{1}{2}  \CC_{4}^{(3)} +  \CC_{9}^{(3)} +\gamma(\bq_1,\bq_2) \left(   \frac{1}{2} \CC_{7}^{(3)}+ \CC_{10}^{(3)} \right) +\beta(\bq_1,\bq_2) \left(   \frac{1}{2} \CC_{6}^{(3)}  + \CC_{11}^{(3)} \right)\bigg]\\
&+ \frac{\bq_{12}\cdot\bq_3}{q_3^2}\left(\frac{1}{2}\CC_4^{(3)} - \CC_9^{(3)}\right) + \frac{\left(\bq_{12}\cdot\bq_3\right)^2}{q_{12}^2q_3^2}\left(\CC_4^{(3)} - \CC_2^{(3)}\right)
\,,
\end{split}
\label{CCfin}
\ee
Using the  structure of the velocity kernel, i.e.,
\be
  \Gtwo(\bq_1,\bq_2 ) = \, 2 \,\beta(\bq_1,\bq_2) +\BC_{1}^{(2)}  \,\gamma(\bq_1,\bq_2)  \;,
  \label{G2fin}
\ee
 gives other five constraints,
\begin{align}
&\frac{1}{2} \CC_{4}^{(3)} + \CC_{9}^{(3)}  = 0, \qquad   \frac{1}{2} \CC_{6}^{(3)} + \CC_{11}^{(3)} = \CC_{0}^{(1)} ,  \quad \frac{1}{2} \CC_{7}^{(3)} + \CC_{10}^{(3)} =  \, \hh \, \CC_{0}^{(1)} \,, \nonumber\\
& \CC_{ 4}^{(3)}  - \CC_{2}^{(3)}  = 0 \,, \qquad  \frac{1}{2} \CC_{4}^{(3)} - \CC_{9}^{(3)}  = 0\,,
\label{C33}
\end{align}
where we have defined
\be
\hh(\eta) \equiv   \int^\eta d\eta'\, f_+(\eta')\left[\frac{D_+(\eta')}{D_+(\eta)}\right]^2 \BC_{1}^{(2)}(\eta') \;.
\ee
Note that the last two conditions in eq.~\re{C33} come directly from the second line of eq.~\re{CCfin}: the structure of these two terms is not present in the velocity kernel, see eq.~\re{G2fin}, due to the momentum conservation for the velocity field,  which constraints the kernel $\Gtwo$ as in eq.~\re{km}.

Only 8 of the above equations are independent, leading to the following relations
\be
\CC_{2}^{(3)}=\CC_{4}^{(3)}=\CC_{9}^{(3)}=\CC_{11}^{(3)}=0\;, \qquad \CC_{3}^{(3)}=  2\, \CC_{0}^{(2)}  \;, \qquad \CC_{6}^{(3)}=2\, \CC_{0}^{(1)}   \;, 
\ee 
and
\be
\label{C7gen}
\CC_{7}^{(3)} = -2\,\CC_{10}^{(3)} + 2 \,\hh \, \CC_{0}^{(1)} \,,  \qquad \CC_{8}^{(3)} = 4\,\CC_{10}^{(3)} + 2\,\CC_{1}^{(2)} - 2 \hh \,\CC_{0}^{(1)} \,,
\ee
leaving with four (i.e.~$\{ \CC_{0}^{(3)}, \CC_{1}^{(3)}, \CC_{5}^{(3)}, \CC_{10}^{(3)} \}$) free coefficients out of 12. The form of the kernel after imposing these relations is given in eq.~\eqref{generalK}.

In the case of a conserved tracer, we  can also impose the two conditions coming from mass and momentum conservation, eq.~\re{km}, which give respectively,
\begin{equation}
\CC_{0}^{(3)}+ \frac{3}{2} \CC_{1}^{(3)} =0,\qquad\qquad \CC_{3}^{(3)}=0\,, 
\label{C34}
\end{equation}
and
\be
\CC_{1}^{(3)} =\CC_{3}^{(3)}=\CC_{9}^{(3)}=0\, , \qquad  \CC_{11}^{(3)}=0\,,
\label{C35}
\ee
allowing us to set 
\be
\CC_{0}^{(3)}=\CC_{1}^{(3)}=\CC_{3}^{(3)}= 0 \;,
\ee
so that only two coefficients are left free. This leads to the matter kernel \eqref{matterK}.

 \section{Time-dependence of the kernels}
 \label{ansols}
 \subsection{$\Lambda$CDM}


In this appendix we give the analytic solutions of the time evolution equations for the coefficients of the third order matter kernels, which are given by
\begin{align}
&(\partial_\chi+2) \AC_{5}^{(3)}(\chi) =  \left( \AC_{1}^{(2)}(\chi)+ \BC_{1}^{(2)}(\chi)\right) -  \AC_{5}^{(3)}(\chi)+ \BC_{5}^{(3)}(\chi)\,,\nonumber\\
&(\partial_\chi +2) \BC_{5}^{(3)}(\chi) = \frac{3}{2} \frac{\Omega_m(\chi)}{f_+^2(\chi)} \left( \AC_{5}^{(3)}(\chi)- \BC_{5}^{(3)}(\chi)\right)\,,\nonumber\\
&(\partial_\chi +2) \AC_{10}^{(3)}(\chi) = -\frac{1}{2} \left( \AC_{1}^{(2)}(\chi)- \BC_{1}^{(2)}(\chi)\right) - \AC_{10}^{(3)}(\chi)+ \BC_{10}^{(3)}(\chi)\,,\nonumber\\
&(\partial_\chi +2) \BC_{10}^{(3)}(\chi) =\frac{3}{2} \frac{\Omega_m(\chi)}{f_+^2(\chi)} \left( \AC_{10}^{(3)}(\chi)- \AC_{10}^{(3)}(\chi)\right)\,.
\label{C3}
\end{align}
The solutions are given by 
\begin{align}
\AC^{(3)}_{5}(\chi)&=\Delta C^{(3)}_{5}(\chi) +\BC^{(3)}_{5}(\chi)=\Delta \AC^{(3)}_{5}(\chi) +\BC^{(3)}_{5}(\chi)\,,\nonumber\\
\BC^{(3)}_{5}(\chi)&=e^{-2\chi} \int_{-\infty}^\chi d\chi' \, e^{2\chi'} \frac{3}{2}\frac{\Omega_m(\chi')}{f^2_+(\chi')}\Delta \AC^{(3)}_{5}(\chi') \,,\nonumber\\
\Delta \AC^{(3)}_{5}(\chi)&=e^{-2\chi} \int_{-\infty}^\chi d\chi' \, e^{2\chi' - \int_{\chi'}^\chi d\chi''\;\left(1+\frac{3}{2}\frac{\Omega_m(\chi'')}{f_+^2(\chi'')} \right)}\, \AC^{(2)}_{+}(\chi')\,,
\end{align}
where 
$\AC^{(2)}_{+}(\chi)\equiv \AC^{(2)}_{1}(\chi)+ \BC^{(2)}_{1}(\chi)$, can be read from eqs.~\re{C21}, \re{C22}, and, analogously, by,
\begin{align}
\AC^{(3)}_{10}(\chi)&=\Delta C^{(3)}_{10}(\chi)+ \BC^{(3)}_{10}(\chi)=\Delta \AC^{(3)}_{10}(\chi)+ \BC^{(3)}_{10}(\chi)\,,\nonumber\\
\BC^{(3)}_{10}(\chi)&=e^{-2\chi}  \int_{-\infty}^\chi d\chi' \, e^{2\chi'}\,\frac{3}{2}\frac{\Omega_m(\chi')}{f^2_+(\chi')}\Delta \AC^{(3)}_{10}(\chi') \,,\nonumber\\
\Delta \AC^{(3)}_{10}(\chi)&=-\frac{ e^{-2\chi}}{2}\, \int_{-\infty}^\chi d\chi' \, e^{2\chi' - \int_{\chi'}^\chi d\chi''\;\left(1+\frac{3}{2}\frac{\Omega_m(\chi'')}{f_+^2(\chi'')} \right)}\AC^{(2)}_{-}(\chi')\,,
\end{align}
where $\AC^{(2)}_{-}(\chi) \equiv \AC^{(2)}_{1}(\chi)- \BC^{(2)}_{1}(\chi)$.

  \subsection{nDGP}
\label{ansolsn}

Here we give the analytic solutions of the time evolution equations for the coefficients of the third order matter kernels in a nDGP cosmology, which are given by
\begin{align}
&(\partial_\chi+2) \AC_{5}^{(3)}(\chi) =  \left( \AC_{1}^{(2)}(\chi)+ \BC_{1}^{(2)}(\chi)\right) -  \AC_{5}^{(3)}(\chi)+ \BC_{5}^{(3)}(\chi) + 2\left(\frac{3}{2}\Omega_m\right)^2 \left(\frac{3}{2}\frac{\Omega_m}{f_+^2}\mu_3 + \frac{\mu_2}{f_+^2} \AC_1^{(2)}\right)\,,\nonumber\\
&(\partial_\chi +2) \BC_{5}^{(3)}(\chi) = \frac{3}{2} \frac{\Omega_m(\chi)}{f_+^2(\chi)} \mu(\chi)\left( \AC_{5}^{(3)}(\chi)- \BC_{5}^{(3)}(\chi)\right)+ 2\left(\frac{3}{2}\Omega_m\right)^2 \left(\frac{3}{2}\frac{\Omega_m}{f_+^2}\mu_3 + \frac{\mu_2}{f_+^2} \AC_1^{(2)}\right)\,,\nonumber\\
&(\partial_\chi +2) \AC_{10}^{(3)}(\chi) = -\frac{1}{2} \left( \AC_{1}^{(2)}(\chi)- \BC_{1}^{(2)}(\chi)\right) - \AC_{10}^{(3)}(\chi)+ \BC_{10}^{(3)}(\chi)\,,\nonumber\\
&(\partial_\chi +2) \BC_{10}^{(3)}(\chi) =\frac{3}{2} \frac{\Omega_m(\chi)}{f_+^2(\chi)} \mu(\chi)\left( \AC_{10}^{(3)}(\chi)- \AC_{10}^{(3)}(\chi)\right)\,.
\label{C3}
\end{align}
The solutions are given by 
\begin{align}
\AC^{(3)}_{5}(\chi)&=\Delta \AC^{(3)}_{5}(\chi)+\BC^{(3)}_{5}(\chi)\,,\nonumber\\
\BC^{(3)}_{5}(\chi)&=e^{-2\chi} \int_{-\infty}^\chi d\chi' \, e^{2\chi'}\,\times\nonumber\\
&\times\,\left[ \frac{3}{2}\frac{\Omega_m(\chi')}{f^2_+(\chi')}\mu(\chi')\Delta \AC^{(3)}_{5}(\chi')+2\left(\frac{3}{2}\Omega_m(\chi')\right)^2 \left(\frac{3}{2}\frac{\Omega_m(\chi')}{f_+^2(\chi')}\mu_3 + \frac{\mu_2(\chi')}{f_+^2(\chi')} \AC_1^{(2)}(\chi')\right) \right]\,,\nonumber\\
\Delta \AC^{(3)}_{5}(\chi)&=e^{-2\chi} \int_{-\infty}^\chi d\chi' \, e^{2\chi' - \int_{\chi'}^\chi d\chi''\;\left(1+\frac{3}{2}\frac{\Omega_m(\chi'')}{f_+^2(\chi'')}\mu(\chi'') \right)}\,\AC^{(2)}_{+}(\chi')\,,
\end{align}
and, analogously, by, 
\begin{align}
\AC^{(3)}_{10}(\chi)&=\Delta \AC^{(3)}_{10}(\chi)+ \BC^{(3)}_{10}(\chi)\,,\nonumber\\
\BC^{(3)}_{10}(\chi)&=e^{-2\chi}  \int_{-\infty}^\chi d\chi' \, e^{2\chi'}\,\frac{3}{2}\frac{\Omega_m(\chi')}{f^2_+(\chi')}\mu(\chi')\Delta \AC^{(3)}_{10}(\chi') \,,\nonumber\\
\Delta \AC^{(3)}_{10}(\chi)&=-\frac{ e^{-2\chi}}{2}\, \int_{-\infty}^\chi d\chi' \, e^{2\chi' - \int_{\chi'}^\chi d\chi''\;\left(1+\frac{3}{2}\frac{\Omega_m(\chi'')}{f_+^2(\chi'')}\mu(\chi'') \right)}\AC^{(2)}_{-}(\chi')\,.
\end{align}

\bibliographystyle{JHEP}
\bibliography{mybibnew.bib}
\end{document}